%% file: conformal-algebra.tex
\documentclass[aps,english,floatfix,
tightenlines,twocolumn,nofootinbib,superscriptaddress]{revtex4-2}

\input{jm-tex-macros-public}
\input{macros}
\usepackage{ragged2e}

\usepackage{array} %
\usepackage{comment}
\usepackage{cleveref}
\usepackage{makecell}
\usepackage{xspace}
\usepackage[normalem]{ulem}

\usepackage[makeroom]{cancel}

\numberwithin{equation}{section}
\renewcommand{\theequation}{\arabic{section}.\arabic{equation}}

\definecolor{XLgreen}{RGB}{34,139,34}
\definecolor{JMblue}{RGB}{25,25,125}
\definecolor{BSorange}{RGB}{140,50,0}
\definecolor{RCpurple}{RGB}{128,0,128}

\newcommand{\JM}[1]{{\color{JMblue}{[JM: #1]}}}

\def\rec{{\text{rec}}}

\begin{document}

\title{Reconstruction of the conformal algebra\\ 
from a critical groundstate on the (fuzzy) sphere}

\author{Rolando Ramirez Camasca}
\affiliation{Department of Physics, University of California at San Diego, La Jolla, CA 92093, USA}
\author{Xiang Li}
\affiliation{Department of Physics, University of California at San Diego, La Jolla, CA 92093, USA}
\affiliation{Mathematical Institute, University of Oxford, Oxford OX2 6GG, UK}
\author{Ting-Chun Lin}
\affiliation{Department of Physics, University of California at San Diego, La Jolla, CA 92093, USA}
\affiliation{Department of Physics, Stanford University, Palo Alto, CA 94305, USA}
\author{John McGreevy}
\affiliation{Department of Physics, University of California at San Diego, La Jolla, CA 92093, USA}

\begin{abstract}
In \cite{Camasca:2026awd} we found a method to reconstruct the dilatation operator (sphere Hamiltonian) from a regulated conformal field theory groundstate.  In this paper, we explore methods to construct the whole global conformal algebra, and test our construction using fuzzy sphere regularization.
Starting from a groundstate of the same Hamiltonian, these methods perform better 
than constructions of the conformal generators based on a Hamiltonian density, 
are model-independent, 
require no additional tuning, 
and improve with system size.  
One purpose of reconstructing the conformal algebra is to identify conformal primary states, which we do with some success. %
We also find further evidence supporting the model of the short-distance corrections to the entanglement Hamiltonian proposed in \cite{Camasca:2026awd} and use it to make systematic improvements.

\end{abstract}

\date{\today}

\maketitle
\let\savedaddcontentsline\addcontentsline
\renewcommand{\addcontentsline}[3]{}

\section{Introduction}
Just as they will find the same sadly-short list of platonic solids in three dimensions as we find, the alien intelligences made from the weird chemistry of a patch of the string landscape with different laws of `fundamental' physics will find the same list of conformal field theories (CFTs).
In this sense, CFTs may be said to be more real than many other objects of interest (e.g.~financial systems, biology, chemistry, or indeed the observable laws of `fundamental' physics).  
Via their relevant deformations, CFTs provide a starting point for understanding all quantum field theories.
They sit in the strategically crucial roads that connect various phases of matter in the kingdom of quantum many-body systems. They govern the universal physics of phase transitions. 

Unfortunately, our means of accessing this reality are limited. 
At least with our current understanding, any concrete description of a nontrivial conformal field theory requires 
some form of truncation to a finite set of data, such as a lattice model, or a subset of basis functions on the space of four-point functions.  (This is why we are sad about the brevity of the list of platonic solids in three dimensions \cite{Brower:2012vg, Brower:2014daa, Lao:2023zis}.)
The fuzzy sphere regularization (\cite{Zhu:2022gjc,Hofmann:2023llr,Hu:2023xak,Zhou:2023qfi,Hu:2024pen,Voinea:2024ryq,Zhou:2024zud,Fardelli:2024qla, Fan:2024vcz,He:2025ong,Zhou:2025rmv,Zhou:2025kng,Guo:2025odn,ArguelloCruz:2025zuq,Taylor:2025odf,Voinea:2025iun,EliasMiro:2025msj,Tang:2025wtj,Fardelli:2026zas,dey2026:O2,dey2026:O3}, for a review see \cite{He:2026ong}) has improved our access to 2+1d CFTs, by providing a slightly-less-crude finitary tool that exactly preserves rotation invariance on the 2-sphere, while maintaining approximate locality. Still, the method so far requires a somewhat arbitrary and artificial choice of critical Hamiltonian.  

By definition, the Hilbert space of a conformal field theory furnishes a representation of the conformal algebra, and so it is a good idea to ask to what extent the regularized Hilbert space is up to this task.  
For the critical Ising model on the fuzzy sphere, \cite{Fardelli:2024qla, Fan:2024vcz, Fardelli:2026zas,Zhou:2026afs} used the artificial critical Hamiltonian density to construct generators of the conformal algebra, and performed many checks (reviewed below) that the low-lying spectrum behaves approximately like that of a CFT representing this algebra, and thus that these generators approximate the conformal algebra.

In this paper, channeling the perspective of Entanglement Bootstrap \cite{Shi:2019mlt, Shi:2020jxd, Shi:2019ngw, Shi:2018krj, Shi:2018bfb, Shi:2020rne, knots-paper, Kim:2021tse, Kim2021, Shi:2023kwr, Kim:2023ydi, Kim:2024amo, Vir, kim2024conformalgeometryentanglement, no-go, figure-eight, Kim:2024gtp, Yang:2025pke, Li:2025sbv, Lin:2023pvl,Li:2025czz, Camasca:2026awd}, we give a model-independent method to extract the conformal algebra directly from the groundstate,
without any further use of the artificial Hamiltonian density, or any further fine-tuning.  The method is not specific to the fuzzy sphere and could be implemented however the CFT is regulated.  
Most of our tests of the method are performed using the fuzzy sphere.  

To convey the idea, we first review the successes of \cite{Camasca:2026awd} at extracting the dilatation operator from the groundstate.  
For a continuum CFT groundstate on $S^d$ of radius $R$, there is a beautiful formula for the entanglement Hamiltonian $K_A = - \log \rho_A$ for a region $A$ that is a round ball of opening angle $\theta_A$: 
\be\label{eq:K-of-ball} K_A^\text{CFT} = \int_{A} d^d\Omega \beta_A(\Omega) T_{00}(\Omega) + f_A \mathbb{I} \ee
where the coolness function is 
$\beta_A(\Omega) = \frac{2\pi R}{\sin\theta_A}(\cos\gamma(\Omega) - \cos\theta_A)$, and $\gamma(\Omega)$ is the angle from the center of $A$ to the point labelled $\Omega$.  
In $d=2$, the c-number 
piece is just the entanglement entropy, $f_A = S_A = { | \partial A | \over \eps} - F $, the area law term minus the RG monotone coefficient $F$.  
This formula follows from the Bisognano-Wichmann (BW) formula \cite{bwtheorem} for the 
entanglement Hamiltonian of the half-space in the groundstate of any relativistic QFT, combined with a conformal transformation that takes the half-space to a round ball \cite{Casini:2011kv, Cardy:2016fqc}.   
If this formula were exactly true of the entanglement Hamiltonian in a {\it regularized} CFT groundstate, we could construct the conformal generators simply by taking appropriate linear combinations of such entanglement Hamiltonians, for example by averaging over the center of $A$. 
In \cite{Camasca:2026awd}, we proposed an ansatz for the entanglement Hamiltonian of a regulated CFT groundstate of the form  
\be K_A = K_A^\text{CFT} + K_A^{\rm UV} ~. \label{truth-vs-reality}\ee
where the error term $ K_A^{\rm UV}$ is 
a sum of local, ultraviolet (UV) contributions 
of short-ranged entanglement along the boundary of $A$, as in \cite{Grover:2011fa}. This ansatz was supported by numerical tests in fuzzy sphere regularization of critical groundstates. 
The equality \eqref{truth-vs-reality} between an operator on the regularized Hilbert space on the left-hand side, and a field theory operator on the right-hand side is intended as a coarse-grained statement, which should hold in the low-energy subspace where the two descriptions overlap.  
We showed that we could construct a good approximation to the dilatation operator 
$ D^{\rm CFT} = R \int_{S^2} d \Omega \,T_{00}(\Omega)$ by choosing a linear combination of concentric balls that cancels the area law term in the entanglement entropy, and averaging their entanglement Hamiltonians over the location of their common center.  

Here we describe an analogous strategy to construct the other conformal generators, beginning with the conformal boost generators $\Lambda_i = P_i + K_i = 2R \int_{S^2} d\Omega \, \hat n_i T_{00}(\Omega)$.
From $\Lambda_i$ and $D$ we can determine 
the momenta $P_i$ and special conformal generators $K_i$ \cite{Fan:2024vcz, Fardelli:2024qla}.

Perhaps surprisingly, when we implement the reconstruction strategy in fuzzy sphere, the operation of averaging over the centers is the bottleneck to studying larger system sizes.  
This is done by projecting onto the subspace of $\gSO(3)$ singlets, which requires decomposing the many-body Hilbert space into $\gSO(3)$ irreps.  Although this operation only needs to be done once, and is a purely group theoretic question, we have not yet found a method which is efficient enough to go beyond $L=15$ spin-half orbitals.  
Since DMRG can find the groundstate for systems that are quite a bit larger, it behooves us to find a way around this obstacle. 

A different reconstruction method that may ameliorate this problem is suggested by the following facts. 
First, for any region $A$ and its complement $\bar A$, 
$K_A - K_{\bar A} $ exactly annihilates the state it was constructed from.  
Via the BW theorem, this purely quantum-information-theoretic fact is deeply related to the boost-invariance of a QFT vacuum.  
Second, for a spherical cap $A$ of opening angle $\theta_A$, the difference of the CFT modular Hamiltonians of $A$ and its complement is
\be K_A^\text{CFT} - K_{\bar A}^\text{CFT} = \frac{2\pi R}{\sin \theta_A} \int  d^d \Omega \( \cos \theta - \cos \theta_A \) T_{00}(\Omega),  \label{eq:full-modular} \ee
where $\theta$ is the angle measured from the center of $A$.  
Third, since we expect the regulated $K_A$ to differ from its CFT counterpart by a sum of local terms at the boundary, we 
expect that the UV contributions in the combination \eqref{eq:full-modular} cancel out. 
Indeed, the existence of this combination 
is even recognized in the algebraic QFT literature 
(unlike all other forms of entanglement Hamiltonian in QFT), 
where 
it is known as the ``full modular Hamiltonian".  
In this context `exist' means `be independent of the UV regulator'.  
So our second method involves linear combinations of ``full modular Hamiltonians'' for different concentric regions.  
A related, further motivation for studying this method is that 
we can expect some cancellation between the local boundary terms in $K_A$ and $K_{\bar A}$. 
A shortcoming of this method is that it necessarily involves regions larger than half the system size, whose reduced density matrix always has a kernel (and many small eigenvalues).  Therefore, this method is more sensitive to the method by which the (effective) kernel of the reduced density matrix is regularized than methods that involve only small subregions.

In principle, these entanglement-based methods require no choice of artificial Hamiltonian at all: we can just begin with a state that is a groundstate of its reconstructed dilatation operator $D_\text{rec}$.  
The problem of searching for all such states on fuzzy spheres of accessible size will be studied elsewhere.  

In the next section, we explain the two strategies, deferring some of the details and a review of the conformal algebra to the appendices.  
In \S\ref{sec:uv-improvements} we use the model of the UV contributions to improve our methods for reconstructing $D$ 
and the other generators.  
In \S\ref{sec:p-and-k} we pause to explain how best to reconstruct $P$ and $K$ once we know $D$ and $\Lambda$.  
In the final three sections, 
we discuss a set of numerical results that address the following three questions: 
1) How faithfully does the reconstruction of conformal generators represent the conformal algebra?   (\S\ref{sec:testing-conformal-alg})
2) What predictions do the reconstructed conformal generators make about the CFT spectra? (\S\ref{sec:extracting-primaries})
3) How can this technology help us search for new CFTs, analogously to \cite{Li:2025czz}? (\S\ref{sec:improved-search})
We present the results of a series of tests of the constructions for the critical Ising model studied in~\cite{Zhu:2022gjc}; the analogous tests for other CFTs that have been regulated on the fuzzy sphere are in the Appendix~\ref{app:other-CFTs}.  

%
%

%
%
%
%
%

%
%
%

%

%
%
%
%
%
%
%
%
%
%
%
%
%
%

%

%
%
%
%

%
%
%

\section{Two strategies for extracting conformal generators}
\label{sec:strategies}

In this section, we describe two different methods of extracting $D_\rec$ and $\Lambda_z^{\rm rec}$ from a regularized CFT groundstate. 
In our exposition here, 
we do not specify how the CFT is regulated.
In our numerical implementations, we mainly use the fuzzy sphere regularization.  
The method can also be applied in lattice regularization, as we describe in Appendix~\ref{app:icosahedron}.

{\bf Averaging method.}  Just as integrating $K_A^\text{CFT}$ of a disk $A$ over the center of the disk, according to the CFT formula \eqref{eq:K-of-ball}, will produce $D$, 
weighting the integral over the center of the disk 
by $\cos\theta$ will produce the conformal boost generator $\Lambda_z = 2R \int \cos \theta T_{00}$.  And just as doing this using the $K_A$ from the regulated groundstate doesn't quite work for $D$ because of the error terms in \eqref{truth-vs-reality}, we need to do something to cancel the UV contributions.  
(More details can be found in Appendix \ref{app:EB-conformal-generators}.)  

To proceed, we take seriously the ansatz \cite{Camasca:2026awd} for the error term 
\be\label{eq:UV-expansion} K_A^{\rm UV} = \int_{\partial A} d\ell\, 
\sum_{j=0} O_j^{\hat n}(\ell) \kappa^j 
~,\ee 
where $\hat n$ is the outward unit normal to the boundary of $A$ within the sphere, $\kappa$ is the extrinsic curvature of the boundary, and $O_\ell^{\hat n}$ are local operators.  

The local boundary expansion is further constrained by purity of the groundstate, as in \cite{Grover:2011fa}. For a pure ground state, the Schmidt decomposition implies
\begin{equation}\label{eq:K_A - K_barA}
    (K_A - K_{\bar A}) | \psi \rangle = 0. 
\end{equation}
Exchanging $A$ with its complement $\bar A$ reverses the orientation of the entangling surface, $\hat n \to - \hat n$, and therefore also the curvature $\kappa \to -\kappa$.
We accordingly organize the local UV expansion into operators of definite parity under reversal of the normal, 
\begin{equation}\label{eq:parity-relation-for-CO}
    \( O_i^{\hat n}(x)  - (-1)^\ell O_\ell^{-\hat n}(x) \)\ket{\psi} = 0 .
\end{equation}

In this paper, we are interested in reconstructing operators out of linear combinations of entanglement Hamiltonians which cancel the short-distance contributions.  For this purpose, it would be very convenient if the operator version of \eqref{eq:parity-relation-for-CO},
\begin{equation}\label{eq:operator-parity-relation-for-CO}
    O_i^{\hat n}(x)  - (-1)^\ell O_\ell^{-\hat n}(x) \buildrel{?}\over{=} 0~,
\end{equation}
were true.  
We should comment that of course in the regulated system, $K_A$ and $K_{\bar A}$ have non-overlapping support and cannot possibly be equal.  
However, on the low-energy subspace,
where we expect \eqref{truth-vs-reality} to hold, the support is smeared and \eqref{eq:operator-parity-relation-for-CO} has a chance to hold.  
We will give some stringent tests of \eqref{eq:operator-parity-relation-for-CO} below and find that it does hold at low energy.

Consider the result of integrating the UV terms \eqref{eq:UV-expansion} against an $\ell=1$ spherical harmonic: 
\bea
\overline K ^{\mathrm{UV}}_{A,(\ell,m)} & \equiv&  \int d\hat\Omega\,Y_{\ell m}^{*}(\Omega) K^{\mathrm{UV}}_{A_{\Omega}}
\\ & = & 
R\sin\theta_{\hspace{-0.05cm}A} \hspace{-0.075cm}\int_{S^2} \hspace{-0.2cm}d\Omega_x \hspace{-0.075cm}\int\hspace{-0.05cm} d\hat n\, Y_{l m}^{*} \!\left(\cos\theta_{\hspace{-0.05cm}A} x- \sin\theta_{\hspace{-0.05cm}A}\hat n\right) \times \nonumber\\ 
   &&  \left(O_0^{\hat n}(x) +\kappa \,O_1^{\hat n}(x)+\cdots\right).
\eea
In the second line, we regarded the integral 
as summing over points $x$ on the sphere, and for each such point $x$, summing over the outward normal vectors $\hat n$ to circles passing through $x$.   Using $\ell=1$ and \eqref{eq:parity-relation-for-CO}, 
\begin{align}
    \overline K^{\mathrm{UV}}_{A,(1,m)} &=R\sin\theta_A\cos\theta_A\,\hspace{-0.075cm}\int_{S^2} \hspace{-0.2cm}d\Omega_x \hspace{-0.075cm}\int\hspace{-0.05cm} d\hat n\,Y_{1m}^{*}( x)O_0^{\hat n}(x) \nonumber \\ &
    - \kappa R\sin^2\theta_{\hspace{-0.05cm}A}\hspace{-0.075cm}\int_{S^2} \hspace{-0.22cm}d\Omega_x \hspace{-0.075cm}\int\hspace{-0.07cm} d\hat n\,Y_{1m}^{*}(\hat n)O_1^{\hat n}(x) +\mathcal{O}(\kappa^2R).
\label{eq:weighted-UV-surviving1}
\end{align}
Since $\kappa = \cot \theta_A/R$, 
successive terms in this expansion are smaller at large $R$, and they have a different dependence on $\theta_A$ than the CFT contributions, and so may be canceled by taking linear combinations of multiple regions of different opening angle.

To reconstruct the whole vector of conformal boost generators $\Lambda_i$, it is useful to define Cartesian-weighted modular Hamiltonians as
\begin{equation}
\overline K_{A,i}
\equiv
2\int_{S^2}d^2\widehat{\Omega}\,
\Omega_i K_{A_\Omega}, \label{eq:cartesian-weighted-modular-Hamiltonian}
\end{equation}
where $\Omega_i$ denotes the Cartesian components of the unit vector specifying the cap center
(for example, the $z$-component is $\overline K_{A,z}=2\sqrt{4\pi/3}\,\overline K_{A,(1,0)}$).  
Using Eq.~\eqref{eq:l1-CFT} and~\eqref{eq:symmetric-generators}, the ideal CFT entanglement Hamiltonians give 
$\overline K_{A,i}^{\mathrm{CFT}}=v_1(\theta_A)\Lambda_i$, 
where $v_1(\theta_A)$ is a weighted volume of region $A$.

The analog of the formula of \cite{Camasca:2026awd} for $D_\text{rec}$, then, is to choose a linear combination of $k=2$ of these objects $\overline K_{A,i} $ to cancel the contribution of $O_0^{\hat n}$ in \eqref{eq:weighted-UV-surviving1}:
\begin{equation}
\Lambda_i^{\mathrm{rec}}
\equiv
\sum_{\alpha=1}^{2}
\widetilde{\lambda}_{A_\alpha}
\overline K_{A_\alpha,i}, 
\end{equation}
with coefficients satisfying
\begin{equation}
\sum_{\alpha}
\widetilde{\lambda}_{A_\alpha}
\sin\theta_{A_\alpha}\cos\theta_{A_\alpha}=0,
\quad
\sum_{\alpha}
\widetilde{\lambda}_{A_\alpha}
v_1(\theta_{A_\alpha})=1. \label{eq:Lambda-k2-conditions}
\end{equation}
The first condition cancels the two leading UV contributions 
in \eqref{eq:weighted-UV-surviving1}, while the second fixes the normalization to the CFT generator.
The fact that the two leading terms in \eqref{eq:weighted-UV-surviving1} both depend on $\theta_A$ as $\cos \theta_A \sin \theta_A$ means that already with only $k=2$ regions we expect the errors to decay with system size.  
In the next section, we describe how to further improve the reconstruction by cancelling the next contribution in this expansion. 

Note that the success of this method, to be shown below, is further evidence for our ansatz for the UV corrections to the regulated entanglement Hamiltonian.  
%
%

{\bf Method with full modular Hamiltonians.}  Alternatively, 
we can use the full modular Hamiltonian, defined as $\widehat K_A  \equiv K_A - K_{\bar A}$, where $\bar A$ denotes the complement of $A$. 
Using~\eqref{eq:full-modular}, we can write the full modular Hamiltonian as 
\be
    \widehat K_A^{\rm CFT} = \frac{\pi }{\sin \theta_A} \( \Lambda_z - 2\cos \theta_A D \).
\ee
We can therefore reconstruct both $D$ and $\Lambda_z$ by taking a linear combination of $k = 2$ concentric regions with different opening angles, 
\be\label{eq:BW-Drec-reconstruction}
    D^{\rm FM}_{\rm rec} = \frac{\sin \theta_{A_1} \widehat K_{A_1} - \sin \theta_{A_2} \widehat K_{A_2}}{2\pi  (\cos \theta_{A_2} - \cos \theta_{A_1})}, 
\ee
and 
\be\label{eq:BW-reconstruction}
    \Lambda_{z, {\rm rec}}^{\rm FM} = \frac{\cos \theta_{A_2} \sin \theta_{A_1} \widehat K_{A_1} - \cos \theta_{A_1} \sin \theta_{A_2} \widehat K_{A_2}}{\pi  (\cos \theta_{A_2} - \cos \theta_{A_1})}.
\ee

Unlike the averaging method, this construction does not require averaging over rotations/decomposing the many-body Hilbert space into irreps of $\gSO(3)$ rotations.
This may therefore allow us to study larger system sizes accessible to DMRG.
In Appendix~\ref{app:BW}, we test this method and discuss its implementation on the fuzzy sphere.

\def\hemis{\mathfrak{h}}
Notice that when $\theta = \pi/2$, the extrinsic curvature $\kappa = 0$. This means that all the UV corrections 
in \eqref{eq:UV-expansion} vanish except the $\ell=0$ term.  
So, from this point of view, $k=1$ region with $A =\hemis $, a hemisphere, is in a certain sense an optimal choice to reconstruct $\Lambda_z$:
\bea K_{\hemis} 
-K_{\overline{\hemis}} &=& K_{\hemis}^{\text{CFT}} 
- K_{\overline{\hemis}}^{\text{CFT}}
+ \oint_{\partial(\hemis)} (O_0^{\hat n} - 
 O_0^{- \hat n} )  \nonumber
 \\ & = & 
 \pi \Lambda_z +  \oint_{\partial\hemis} (O_0^{\hat n} - 
 O_0^{- \hat n} ) ,  \label{eq:test-operator-equation}\eea 
 or at least to test our model of the UV contributions.
Any remaining error is a shortcoming of either 
\eqref{eq:parity-relation-for-CO} or for 
\eqref{eq:KUV-local} as an expression for $K_A - K_A^\text{CFT}$.
In Appendix \ref{app:BW} we use \eqref{eq:test-operator-equation} to test \eqref{eq:operator-parity-relation-for-CO}.
We note that in addition to involving the dangers of small and zero eigenvalues of the reduced density matrix of large regions, the choice of $A = $ the hemisphere has the severe conceptual drawback that it is not local --  the symmetry charge reconstructed using this choice is not a sum of local operators.

\section{Canceling UV contributions}
\label{sec:uv-improvements}

One feature we observed in \cite{Camasca:2026awd}, but did not understand clearly, is the noticeable dependence of the results on the choice of subregions used in the reconstruction of $D_\text{rec}$. For example, at fixed system size, the cost function measuring similarity to the CFT spectrum 
\cite{Fan:2024vcz,Fan:2025bhc} improves as the subregions become larger. In that paper we chose subregion sizes empirically and studied their behavior as the system size increased. 
Here we pause in our construction of conformal generators to explain this subregion dependence within the UV model we proposed, \eqref{eq:KUV-local}.
We will then use this understanding to improve the construction of both $D_\text{rec}$ as well as that of the conformal generators.  

To obtain $D_\text{rec}$, we took a linear combination of $\gSO(3)$-averaged entanglement Hamiltonians associated with $k$ regions:
\be
D_\text{rec} = \sum_{\alpha=1}^k \lambda_{A_\alpha} \bar K_{A_\alpha},
\ee
where $\bar K_{A_\alpha}$ is the averaged modular Hamiltonian of region $A_\alpha$. Using $k = 2$ regions, we impose two conditions: (i) cancel the area law: $\sum_\alpha \lambda_{A_\alpha} \sin \theta_{A_\alpha} = 0$, and (ii) normalize the CFT dilatation operator: $\sum_\alpha \lambda_{A_\alpha} V_{A_\alpha} = 1$. 
Here $V_{A} \equiv 
\frac{R}{\mathrm{Vol}(S^2)}\int d^2 \Omega \beta_A(\Omega) =  2 \pi\sin^4(\theta_A/2)/\sin(\theta_A)$. 
Therefore, we found something of the form
\be 
D_\text{rec} = D^\text{CFT} + \text{subleading UV terms of order $1/R$}. \label{eq:Drec-subleading-UV}
\ee
In the fuzzy sphere regularization, the radius of the sphere $R = \sqrt{L}$ where $L = 2s+1$ is the number of orbitals. Therefore, corrections of order $1/R$ scale as $1/\sqrt{L}$.
We claim that these subleading terms have all the subregion dependence, and that this dependence can be made completely explicit. This means that we can cancel it.  
\begin{figure}[!t]
    \centering
    \includegraphics[width=\linewidth]{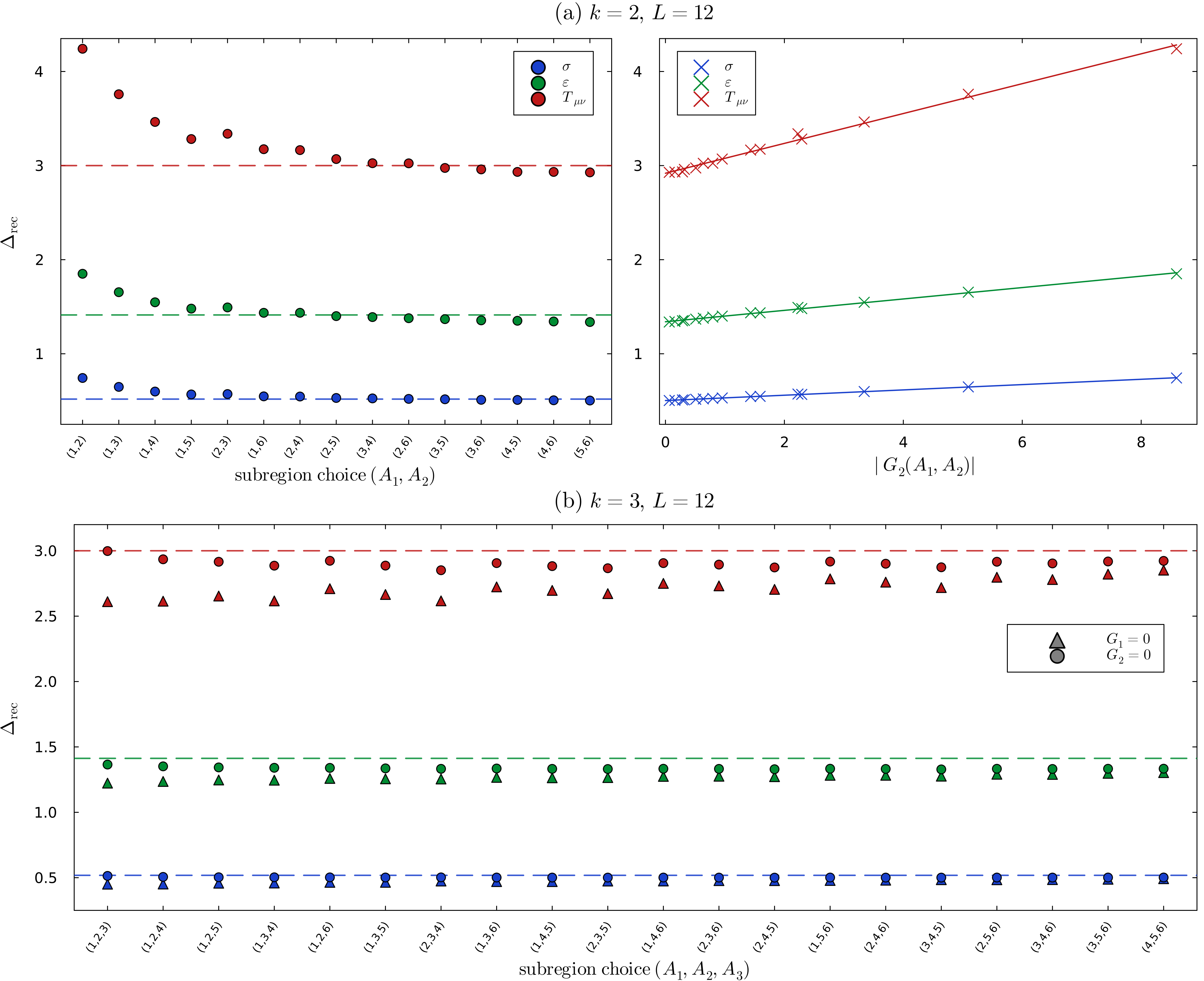}
    \caption{\justifying Low-energy primaries of $D_{\rm rec}$ obtained from the critical Ising model at $L=12$ as a function of the subregion choice. 
A subregion is specified by a number of adjacent orbitals including an orbital with extremal $L_z$.  The opening angle $\theta_A$ is determined by assuming that the ratio of the number of orbitals in region $A$ and the total number of orbitals $L$ is equal to the ratio of the area $A$ and the total area of the sphere: $\theta_{A_i} = \arccos(1-2|A_i|/L)$. 
    Horizontal dashed lines denote the conformal bootstrap (CB) values of the dimensions.
    (a) Reconstructed spectrum using $k = 2$ regions. 
    Left: Subregion choices are ordered by decreasing value of $|G_{2}(\{A_i\})|$.
    Deviations are largest for small subregions and decrease as the regions become larger, consistent with the UV model.
    Right: As a function of $|G_{2}(A_1,A_2)|$. 
    The low-energy spectrum follows a linear behavior, in agreement with the leading UV term coming from the $G_2$ term.
    (b) Reconstructed spectrum using $k=3$ regions.
    Triangles and circles correspond to the additional condition being $G_{1} = 0$ and $G_{2}= 0$ respectively.
    Cancelling the $G_2 = 0$ contribution strongly suppresses the dependence on the subregion choice and improves the agreement with the CB values. }
    \label{fig:UV-improvements}
\end{figure}

More explicitly, consider the next subleading UV terms coming from the curvature expansion:
\bea 
D_\text{rec} &=& D^\text{CFT} + \sum_\alpha \lambda_{A_\alpha} \hspace{-0.075cm}\sin \theta_{A_\alpha} \hspace{-0.075cm}
\( \sum_{\ell = 1} \kappa_{A_\alpha}^\ell \hspace{-0.1cm}\int \hspace{-0.1cm}dx \hspace{-0.1cm}\int \hspace{-0.1cm}Rd\hat n \,O_\ell^{\hat n}(x) \) 
\nonumber
\\ & \equiv&  
D^\text{CFT} + 
\sum_{\ell = 1} G_\ell(\{A_\alpha\})/R^{\ell-1} \hspace{-0.075cm} 
\int \hspace{-0.1cm}dx \hspace{-0.1cm}\int \hspace{-0.1cm}d\hat n \,O_\ell^{\hat n}(x) 
\label{eq:def-of-Gell}
\eea
where the subregion dependence of the subleading terms is in $G_\ell({A_\alpha})$. 
First, the contributions from odd $\ell$ vanish 
by the fact that $ O^{-n}_\ell = (-1)^\ell O^n_\ell$.  
Second, the curvature $\kappa_A$ for a circular boundary on the sphere is given by $\kappa_A = \cot \theta_A/R$. Therefore, the leading correction is proportional to
\be 
G_2(\{A_\alpha\}) = \sum_{\alpha=1}^k \lambda_{A_\alpha} \sin \theta_{A_\alpha} \cot^2 \theta_{A_\alpha}.
\ee
This explains a fact we had observed numerically in \cite{Camasca:2026awd}: 
Smaller subregion choices give a spectrum farther from the CFT: $\cot \theta$ for small $\theta$ becomes large and so the subleading terms are enhanced. 
And it explains how to improve the result by taking $k=3$ regions: 
the third condition we should impose is $0 \buildrel{!}\over{=} G_2 = \sum_\alpha \lambda_{A_\alpha} \sin \theta_{A_\alpha} \cot^2 \theta_{A_\alpha}$.
Fig.~\ref{fig:UV-improvements} shows the dependence on the subsystem size of the low-lying spectrum of $D_\text{rec}$ with and without the imposition of this condition. 
The inference we make from Fig.~\ref{fig:UV-improvements} is that imposing $G_1=0$ doesn't help because the quantity that $G_1$ multiplies in \eqref{eq:def-of-Gell} is already zero;
this quantity is the one in \eqref{eq:operator-parity-relation-for-CO}.  Thus, these results further support the model of the UV contributions to the entanglement Hamiltonian we are using, and offer an improvement over previous work \cite{Camasca:2026awd}.

The analogous improvement for the conformal boosts follows from the weighted average introduced in \S\ref{sec:strategies}. 
Within our UV model, the first condition in Eq.~\eqref{eq:Lambda-k2-conditions} cancels the two leading terms in the curvature expansion.
We can further improve the reconstruction by taking a linear combination of $k=3$ regions,
\begin{equation}
\Lambda_i^{\mathrm{rec}}
\equiv
\sum_{\alpha=1}^3
\widetilde{\lambda}_{A_\alpha}
\overline K_{A_\alpha,i}, 
\end{equation}
and imposing the additional condition 
\be
\sum_\alpha \tilde \lambda_{A_\alpha}
\sin\theta_{A_\alpha}\cos\theta_{A_\alpha} \cot^2 \theta_{A_\alpha}=0,
 \ee
which cancels the next subleading UV contributions 
in \eqref{eq:weighted-UV-surviving1}, as derived in Appendix~\ref{app:EB-conformal-generators}.  
Fig.~\ref{fig:Lambda-rec-comparison-k} in the Appendix quantifies the improvement accomplished by imposing this condition.  
This is the prescription we use to reconstruct $\Lambda_i$ in the rest of the paper.

\section{$P$ and $K$ from $\Lambda$ and $D$}
\label{sec:p-and-k}
\begin{figure}[!t]
    \centering
    \includegraphics[width=\linewidth]{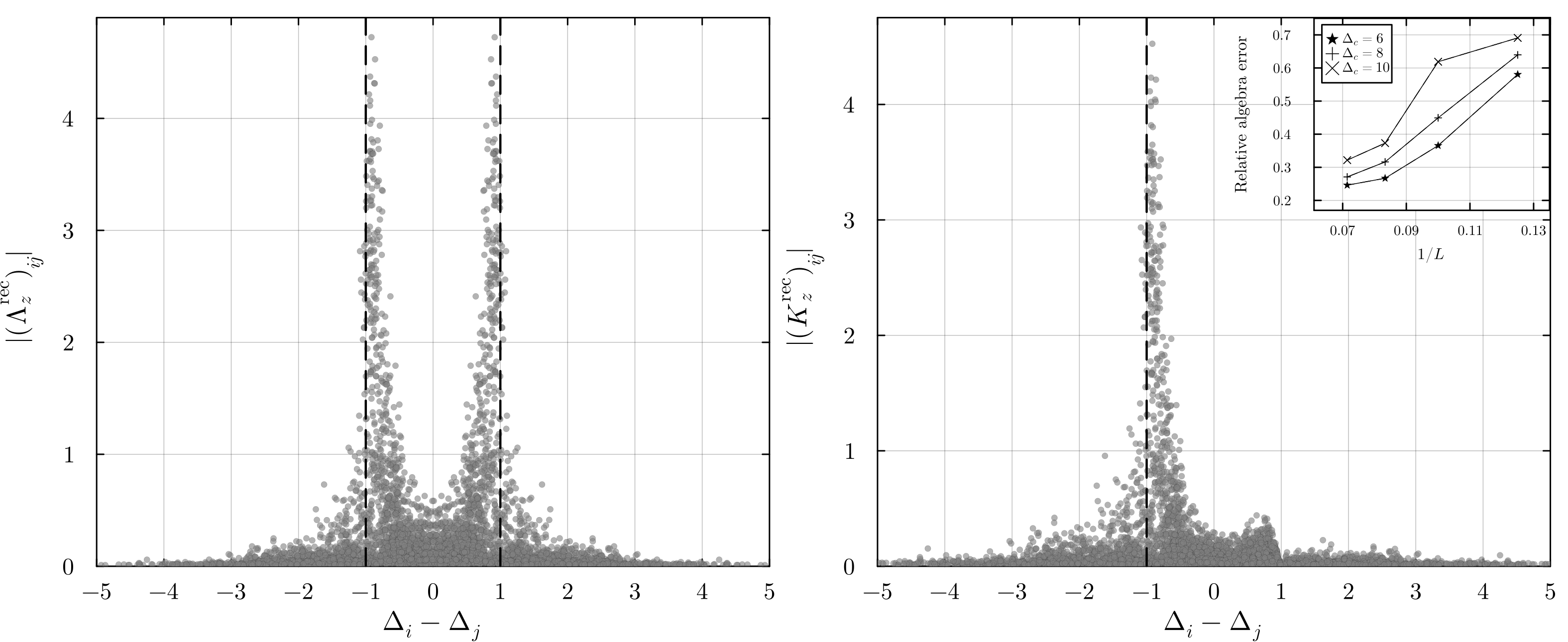}
    \caption{\justifying Matrix elements of the reconstructed conformal boost $\Lambda_z^{\rm rec}$~(left) and lowering operator $K_z^{\rm rec}$~(right) in the $D_{\rm rec}$ eigenstates as a function of the energy difference $\Delta_i - \Delta_j$. 
    Here, we consider all states with energy $\Delta <8$ of the critical Ising model at $L=14$ with subregions $(A_1,A_2,A_3)=(1,4,6)$  and plot matrix elements bigger than $10^{-2}$.
    Dashed vertical lines denote the exact CFT expectations. 
    Inset:
    Relative Frobenius norm error in the commutation relation $[D_{\rm rec}, K_z^{\rm rec}] = -K_z^{\rm rec}$ evaluated within the low-energy spectrum $\Delta < \Delta_c$ as a function of inverse system size, for $\Delta_c = 6,8,10$.
    We see that the error decreases with increasing system size.}
    \label{fig:CFT-matrix-elements-energy}
\end{figure}
Before discussing the numerical results, we need to specify how we construct the translation $P_i$ and special conformal $K_i$ generators from the dilatation operator $D$ and the conformal boost $\Lambda_i$. 
Previous works used different prescriptions. In~\cite{Fardelli:2024qla}, the authors used the fact that, in the ideal CFT, $P_z$ and $K_z$ raise and lower the eigenvalues of $D$ by exactly one.
Hence, the generators $P_z$ $(K_z)$ were obtained by selecting the entries of $(\Lambda_z)_{ij}$ in the $D$ eigenbasis with $i>j$ $(i<j)$.
In contrast,~\cite{Fan:2024vcz,Zhou:2026afs} use the algebraic relations 
\be \label{eq:ruihua-formula-for-PK} 
    P_z =  \frac{1}{2}(\Lambda_z +  [D, \Lambda_z]), ~~~
    K_z  =  \frac{1}{2}(\Lambda_z -  [D, \Lambda_z]). 
\ee
In the eigenbasis of $D$, this gives
\be
\begin{split}
    (P_z)_{ij} &= \frac{1 +  (\Delta_i - \Delta_j)}{2} (\Lambda_z)_{ij}, \\
    (K_z)_{ij} &= \frac{1 -  (\Delta_i - \Delta_j)}{2} (\Lambda_z)_{ij}.
\end{split}
\ee
The two prescriptions are equivalent in the exact CFT, where the matrix elements of $\Lambda_z$ are only nonzero for $\Delta_i - \Delta_j = \pm 1$. 
However, they differ at finite system size where the generators do not satisfy the exact CFT selection rules.
While the first prescription separates the matrix elements according to whether they raise or lower the energy, the second weights them according to their actual energy differences.
In the following, we adopt the latter prescription, replacing $D$ and $\Lambda_z$ with their reconstructed counterparts, which we find to give better results at a fixed system size. Moreover, we construct the conformal generators on the full finite-size Hilbert space, without restricting them to a low-energy subspace. (When measuring violations of conformal relations (like $K_z \ket{\text{primary}} = 0$) by the norms of states, restricting to a low-energy subspace (as in \cite{Fardelli:2024qla, Fardelli:2026zas, Zhou:2026afs} but not \cite{Fan:2024vcz}) gives only a lower bound on the error.)
 
We first examine the matrix elements of the reconstructed conformal generators. Figure~\ref{fig:CFT-matrix-elements-energy} shows the reconstructed conformal boost $\Lambda_z^{\rm rec}$ and the lowering operator $K_z^{\rm rec}$ in the $D_{\rm rec}$ eigenbasis as a function of the energy difference $\Delta_i - \Delta_j$.
In the CFT limit, the matrix elements of $\Lambda_z = P_z + K_z$ are nonzero only between states whose energies differ by one, while $K_z$ only connects states differing by minus one. 
We see that the matrix elements of $\Lambda_z^{\rm rec}$ are concentrated around the expected values $\Delta_i - \Delta_j = \pm 1$. 
This is nontrivial since these selection rules were not imposed in extracting $\Lambda_z^{\rm rec}$ from the ground state nor was the normalization fixed using additional CFT data.
The commutator prescription then produces the conformal generator $K_z^{\rm rec}$ whose matrix elements are concentrated near $\Delta_i - \Delta_j = - 1$.
In the Appendix \ref{app:icosahedron} we show the same tests for the regularization of the critical Ising model using the icosahedron, which works surprisingly well.

\begin{figure}[!t]
    \centering
    \includegraphics[width=\linewidth]{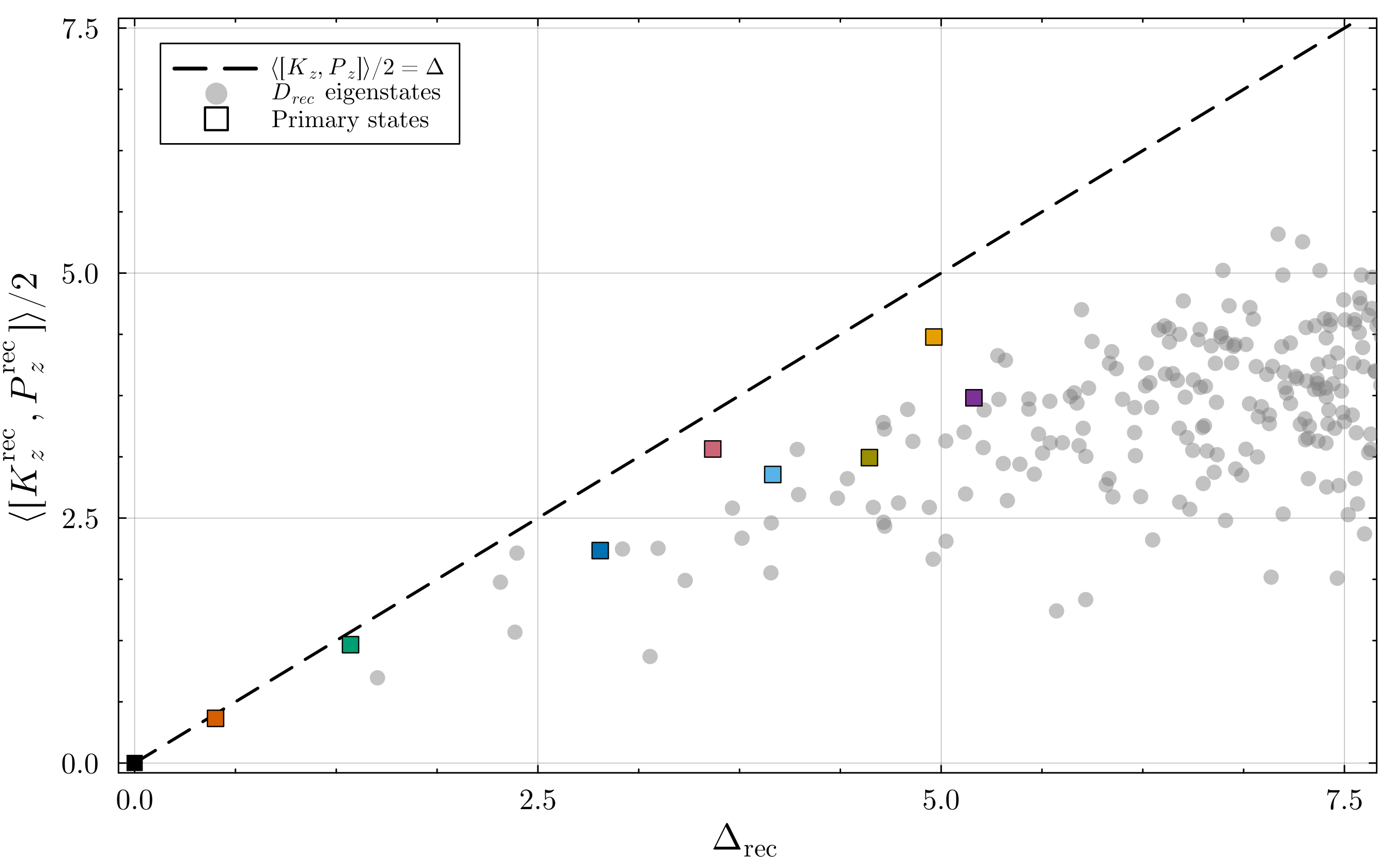}
    \caption{\justifying Expectation value of the commutator $\tfrac{1}{2}[K_z^{\rm rec},P_z^{\rm rec}]$ for eigenstates of $D_{\rec}$ with angular momentum $\ell \leq 5$, as a function of their energy $\Delta_{\rec}$.
    Here, we fix the system size $L=14$ and subregions $(A_1,A_2,A_3) = (1,4,6)$.
    Low-energy primaries are plotted using squares.
    In the CFT limit, all states should lie on the dashed diagonal line.
    }
    \label{fig:CFT-algebra-Drec}
\end{figure}

\section{Testing the conformal algebra}\label{sec:testing-conformal-alg}
\begin{figure}[!t]
    \centering
    \includegraphics[width=\linewidth]{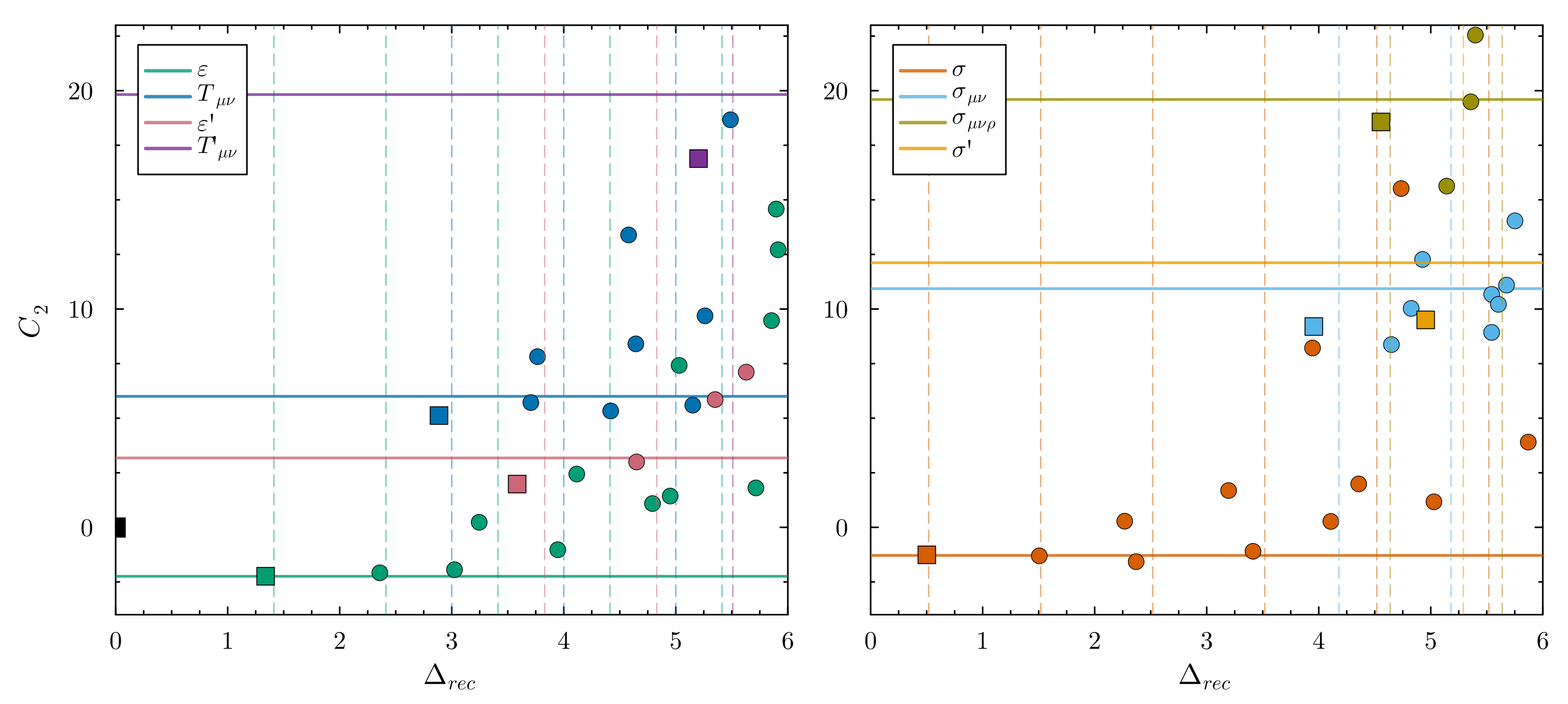}
    \caption{\justifying Expectation value of the conformal Casimir $C_2$ for eigenstates of $D_{\rm rec}$ at $L=14$ and subregion $(A_1,A_2,A_3) = (1,4,6)$. 
    Left and right subplots show the $\mathbb{Z}_2$-even and $\mathbb{Z}_2$-odd sectors respectively. 
    Squares denote the primary states, while circles correspond to the descendants identified by acting with the raising operator $P_i^{\rm rec}$ onto the primaries.}
    \label{fig:CFT-Casimir-Drec}
\end{figure}

Although the reconstructed generators exhibit the expected raising and lowering behavior, this does not guarantee that they satisfy the full conformal algebra. 
We therefore perform two additional tests as in~\cite{Fardelli:2024qla}.
First, we examine whether the reconstructed generators satisfy the CFT commutation relation $[K_z, P_z] = 2D$~(See Appendix~\ref{app:CFT-basics} for a review).
Figure~\ref{fig:CFT-algebra-Drec} shows its expectation value in the $D_{\rm rec}$ eigenstates as a function of their reconstructed energies.
In contrast to the $L=14$ calculation in~\cite{Fardelli:2024qla}, we evaluate the commutator without truncating the intermediate energy eigenstates.
Second, we check whether states within the same conformal family have the same value for the conformal Casimir $C_2 = D (D-3)  + L^2-  \sum_i P_i K_i$. 
Figure~\ref{fig:CFT-Casimir-Drec} shows the Casimir expectation values for primaries and their descendants, identified using the reconstructed raising operators $P_i^{\rm rec}$ on the primaries.
We find that the low-energy states approximately satisfy the commutation relation and reproduce the expected Casimir values for different conformal families, while deviations become more pronounced at higher energies. 

\begin{figure}[!ht]
    \centering
    \includegraphics[width=\linewidth]{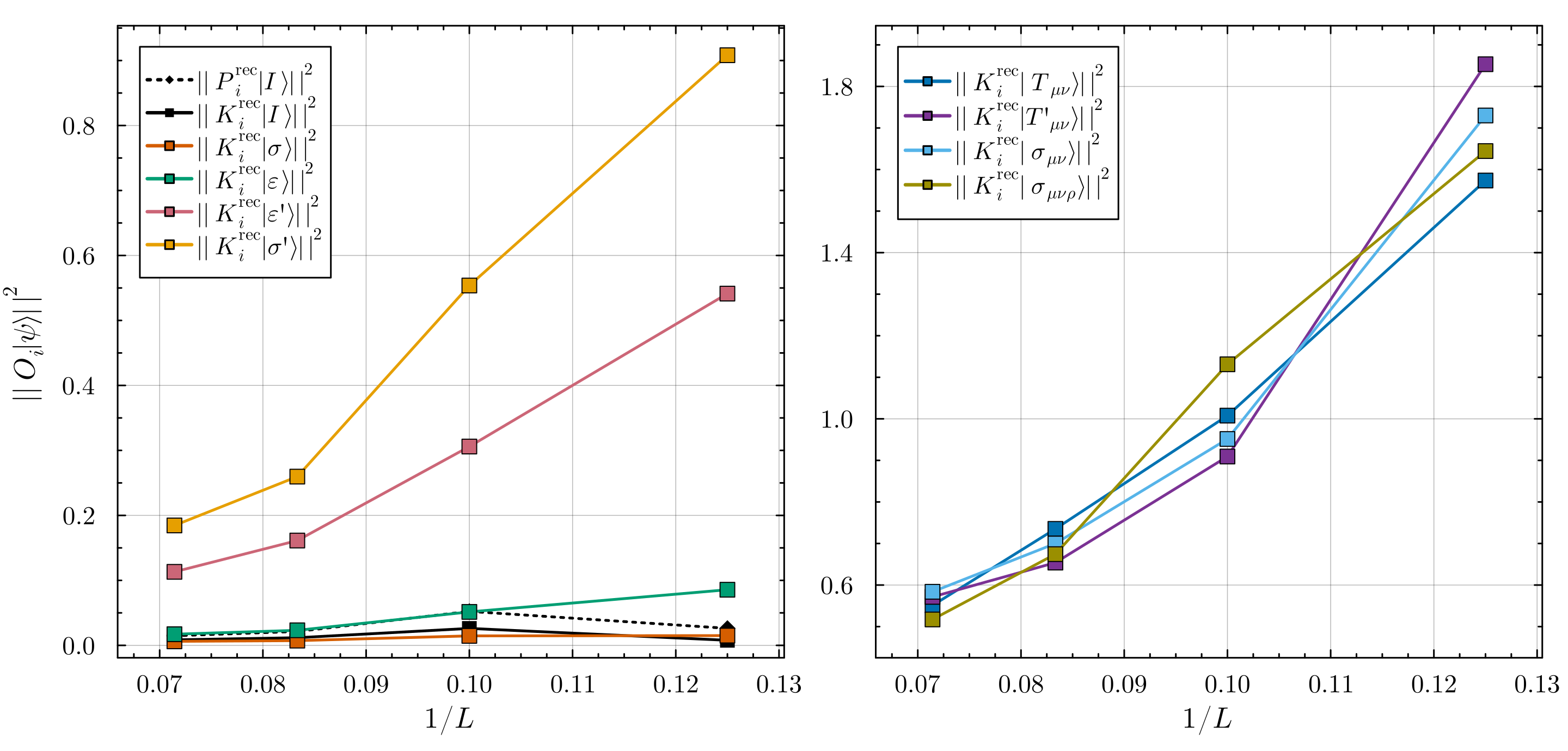}
    \caption{\justifying Sum of the squared norms of the reconstructed generators acting on the vacuum and low-energy primary states of $D_{\rec}$ as a function of inverse system size. 
    Here, we are choosing the subregions to scale with system size as $(A_1,A_2,A_3) = (\left\lfloor L/10 \right \rceil, \left\lfloor L/4 \right\rceil, \left\lfloor 2L/5 \right\rceil)$, where $\left\lfloor \ldots \right\rceil$ denotes the nearest integer. 
    Left subplot corresponds to scalar primaries while the right subplot to spinning primaries.
    }
    \label{fig:Norm-generators-L}
\end{figure}

Finally, we examine whether the action of the reconstructed generators on primary states improves with increasing system size. 
In the exact CFT, the lowering generators $K_i$ annihilate all primary states while both $P_i$ and $K_i$ annihilate the vacuum.
Figure~\ref{fig:Norm-generators-L} shows the corresponding squared norms, summed over all spatial directions, as a function of the inverse system size, for several scalar and spinning primaries. 
We observe that these norms decrease with system size, with larger deviations for spinning primaries. 
For comparison, Ref.~\cite{Fan:2024vcz} found that the squared norms increase with system size,
and are bigger by an order of magnitude.

In Appendix \ref{app:BW}, 
we also test the reconstruction of the conformal algebra generators using the `full modular Hamiltonian' described in \eqref{eq:BW-reconstruction}. 
We find reasonable agreement within the low-energy spectrum. %

\section{Extracting primaries from the $D_{\rm rec}$ spectrum}\label{sec:extracting-primaries}
\begin{figure}[!t]
    \centering
    \includegraphics[width=\linewidth]{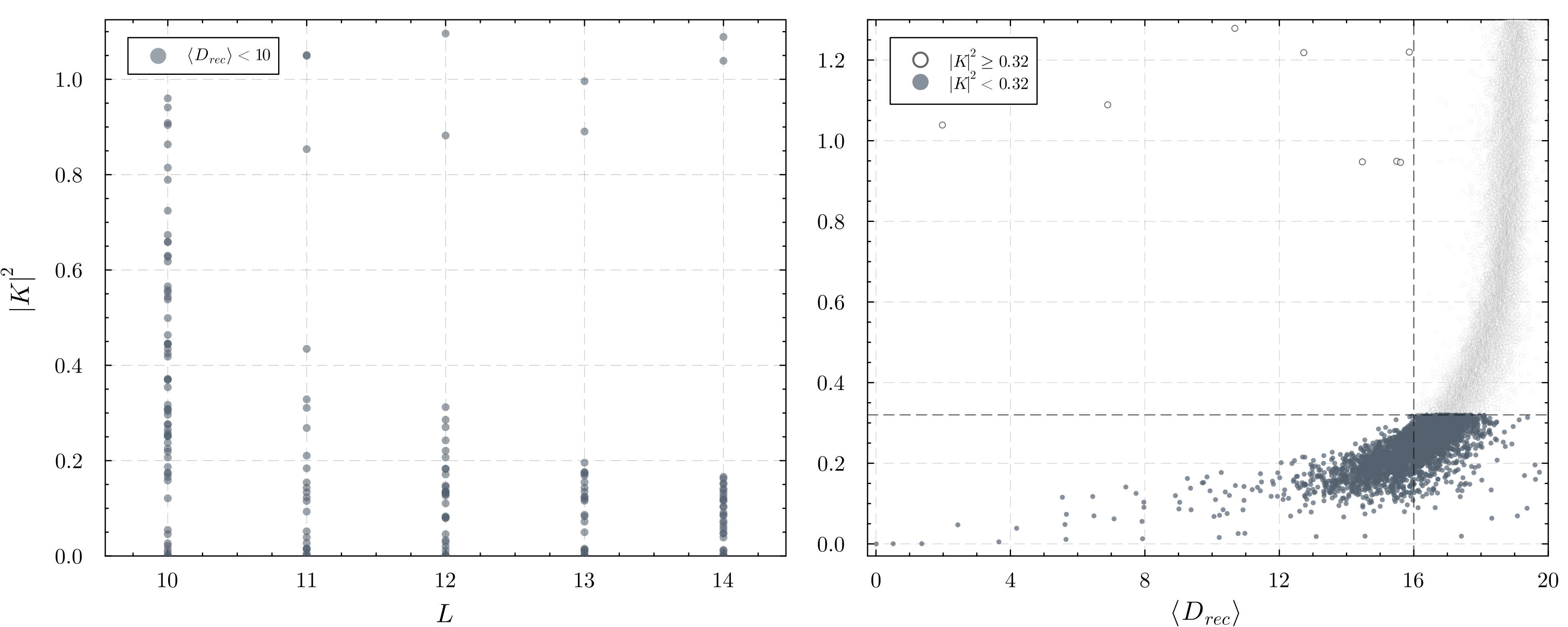}
    \caption{\justifying Low-lying eigenvalues of $|K|^2 = \sum_i (K_i^{\rm rec})^\dagger K^{\rm rec}_i$ for the critical Ising model.
    Here, the subregions scale with system size as $(A_1,A_2,A_3) = (\left\lfloor L/10 \right \rceil, \left\lfloor L/4 \right\rceil, \left\lfloor 2L/5 \right\rceil)$, where $\left\lfloor \ldots \right\rceil$ denotes the nearest integer.
    Left: Dependence on system size $L=10,\ldots,14$, for states $\langle D_{\rm rec} \rangle \leq 10$ with angular momentum $\ell \leq 5$.  
    Right: Eigenvalues of $|K|^2$ as function of $\langle D_{\rm rec} \rangle$ at $L=14$.
    }
    \label{fig:gap-opening-L}
\end{figure}

Having tested the reconstructed conformal algebra, we now investigate whether the generators can be used to 
identify primary states in the CFT spectrum.
Following the approach proposed in~\cite{Fardelli:2026zas}, we construct the operator: 
\be
    |K|^2 \equiv \sum_{i} (K_i^{\rm rec})^\dagger K_i^{\rm rec},
\ee
whose kernel in the CFT limit comprises primary states. 
Apart from a few special descendants, such as $\square \sigma$, generic descendants in a unitary CFT have $|K|^2\geq 1$.
This gap provides a way to distinguish primaries from descendants by looking for states with small $|K|^2$. 
Figure~\ref{fig:gap-opening-L} shows the low-lying eigenvalues of $|K|^2$ as a function of system size and reconstructed energy. 
We observe that a set of low-lying states separates from the rest of the spectrum as the system size increases.
At fixed system size, this gap is more pronounced at low-energies while it becomes less clear for higher energies.

\begin{figure}[!ht]
    \centering
    \includegraphics[width=\linewidth]{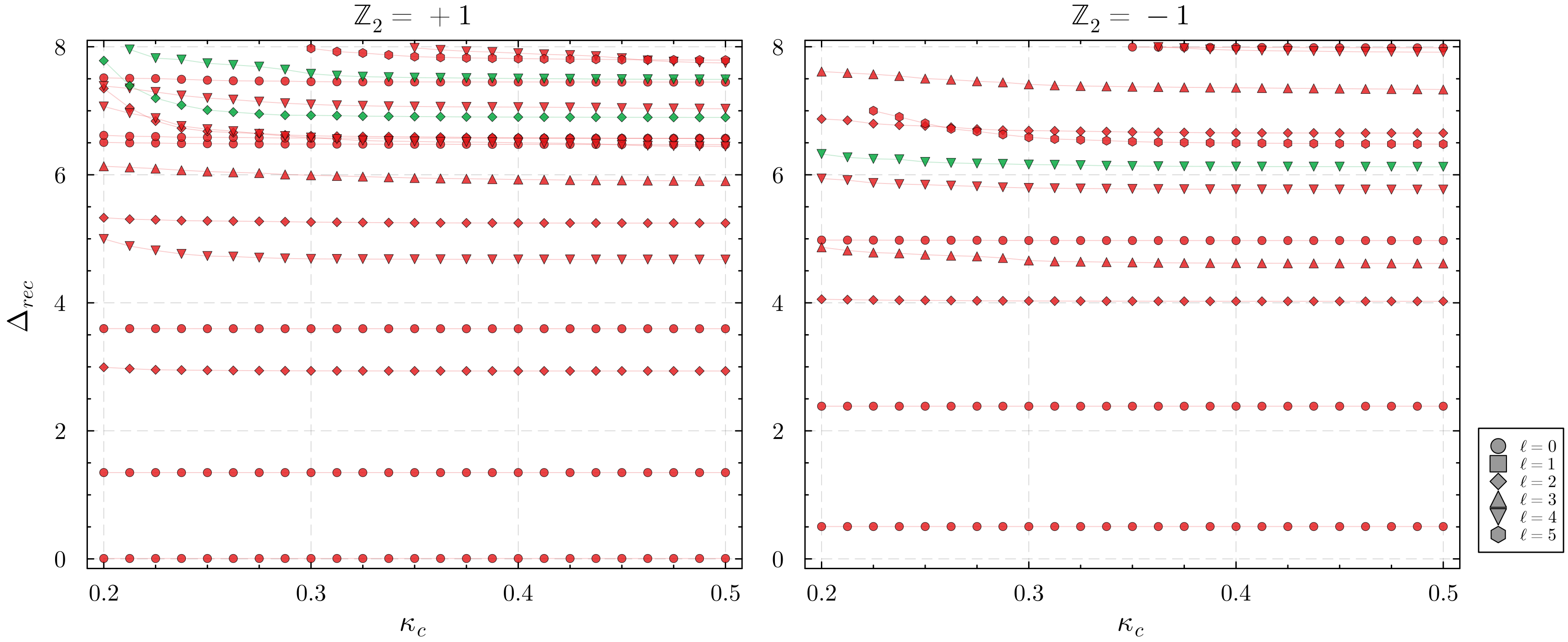}
    \caption{\justifying Dependence of the reconstructed spectrum on the $|K|^2$ cutoff $\kappa_c$ in the critical Ising model at $L=14$ with subregions $(A_1,A_2,A_3)=(1,4,6)$. 
    Scaling dimensions are obtained by rediagonalizing $D_{\rm rec}$ within the subspace of $|K|^2$ eigenstates with eigenvalues below $\kappa_c$. 
    Left and right subplots correspond to the $\mathbb{Z}_2$ even and odd sectors, while red and green to particle-hole even and odd states respectively. Marker shape specifies angular momentum $\ell$.}
    \label{fig:stability-low-energy-spectrum-cutoff}
\end{figure}

We therefore define a subspace of candidate primaries by selecting eigenstates with $|K|^2$ below a cutoff $\kappa_c$.
Since the eigenstates of $|K|^2$ are not necessarily eigenstates of $D_{\rm rec}$ at finite size, we rediagonalize $D_{\rm rec}$ within this subspace to obtain the reconstructed scaling dimensions.
However, there is some arbitrariness in the choice of $\kappa_c$.
Hence, we study how the resulting scaling dimensions depend on this cutoff.
Figure~\ref{fig:stability-low-energy-spectrum-cutoff} shows that the low-energy spectrum is not very sensitive to the choice of $\kappa_c$.   
We note further that the range of $\kappa_c$ below which we capture the known low-lying primaries is an order of magnitude smaller than that required in \cite{Zhou:2026afs}. 

\begin{figure}[h]
    \centering
    \includegraphics[width=\linewidth]{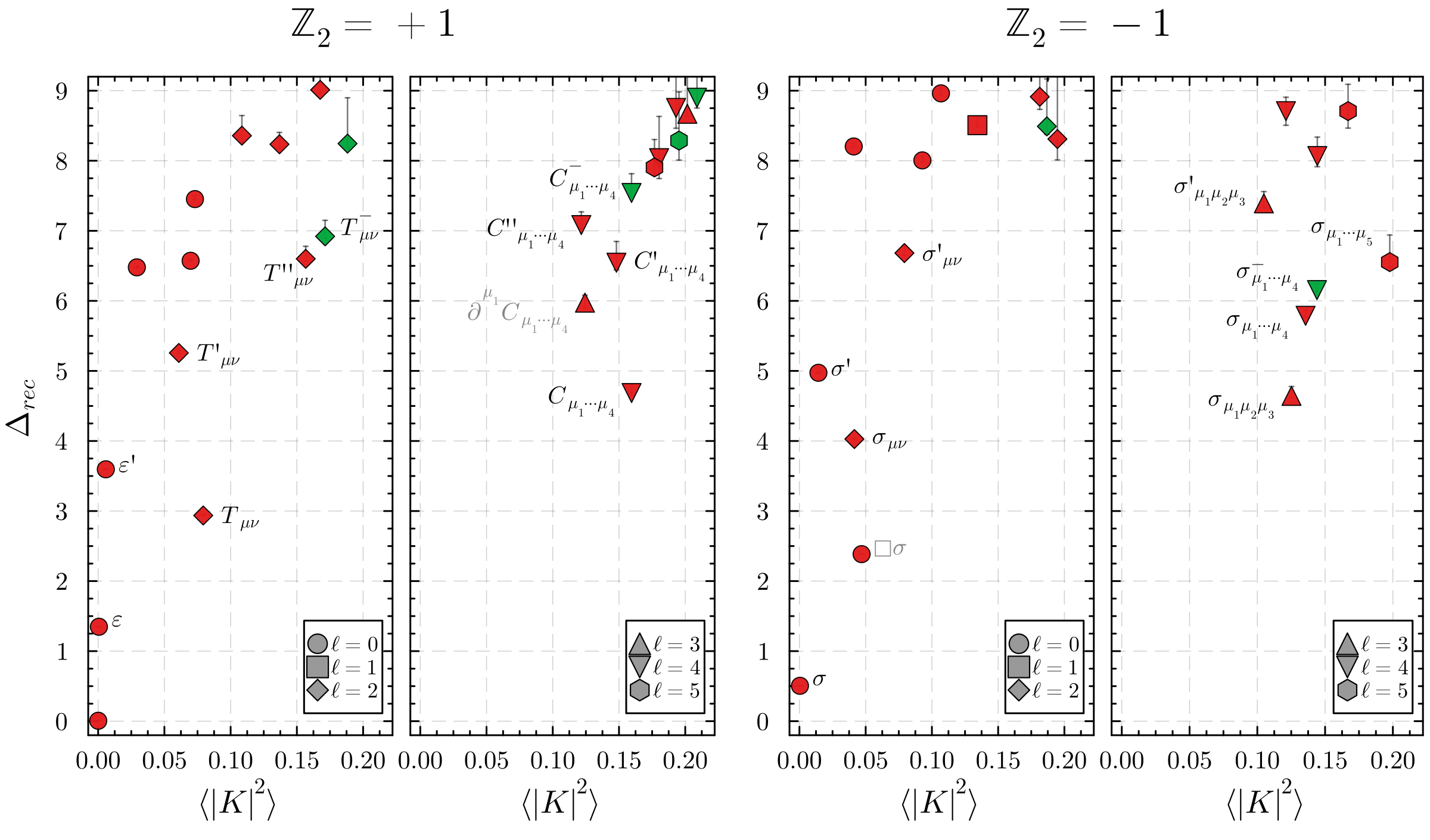}
    \caption{\justifying Reconstructed low-$|K|^2$ spectrum of the critical Ising model at $L=14$ with subregions $(A_1,A_2,A_3)=(1,4,6)$. 
    We select the eigenstates of $|K|^2$ with eigenvalues below $\kappa_c = 0.32$, shown in Figure~\ref{fig:gap-opening-L}, and rediagonalize $D_{\rm rec}$ within this subspace.
    The resulting reconstructed spectra are plotted against $\langle |K|^2 \rangle$, separated into the $\mathbb{Z}_2$-even (left two plots) and $\mathbb{Z}_2$-odd (right two plots) sectors, and, for clarity, into low and high angular momentum sectors.
    Red and green colors denote particle-hole even and odd states respectively, while marker shape specifies angular momentum $\ell = 0, \ldots, 5$.
    We label some low-energy primary candidates and small $|K|^2$ special descendants. 
    Vertical error bars represent the variation of the reconstructed $\Delta_{\rm rec}$ under changes of the cutoff over $\kappa_c \in [0.22, 0.5]$ (and are not an estimate of finite-size error). 
    }
    \label{fig:Low-K2-primaries-L14}
\end{figure}

Finally, Figure~\ref{fig:Low-K2-primaries-L14} shows the reconstructed low-$|K|^2$ spectrum of the critical Ising model at our largest system size available $L=14$, separated into the $\mathbb{Z}_2$ even and odd sectors.
Each sector is further divided into states with angular momentum $\ell=0,1,2$ and $\ell=3,4,5$ for clarity. 
Vertical error bars represent the variation of the reconstructed scaling dimensions under changes in the cutoff $\kappa_c$. 
We label selected states according to their expected CFT counterparts, based on their symmetry sector and reconstructed energy.
We recover the well-known primaries $\sigma,\, \varepsilon,\, T_{\mu \nu}, \varepsilon', \,\sigma_{\mu \nu}$ and $\sigma_{\mu_1 \mu_2 \mu_3}$, previously identified from the spectrum in~\cite{Camasca:2026awd}.
We also find the expected low-$|K|^2$ descendant states $\square \sigma$ and $\partial^{\mu_1} C_{\mu_1 \ldots \mu_4}$ as discussed in~\cite{Fardelli:2026zas}, where $C_{\mu_1 \ldots \mu_4}$ is the leading $\mathbb{Z}_2$ even, particle-hole even, $\ell = 4$ primary. 

Beyond these familiar states, our results show good agreement with the primary spectra reported in~\cite{Fardelli:2026zas, Zhou:2026afs}, as well as with predictions from the conformal bootstrap and the $\varepsilon$-expansion~\cite{Simmons:2016wlq,Kos:2016ysd,Chang:2024whx, Henriksson:2022gpa,Henriksson:2022rnm}.
In particular, we find states consistent with the particle-hole odd primaries $\sigma^-_{\mu_1\ldots\mu_4}$, $T^-_{\mu\nu}$, and $C^-_{\mu_1\ldots\mu_4}$ identified in~\cite{Fardelli:2026zas}. 
Their reconstructed dimensions are approximately $6.15$, $6.92$, and $7.54$, about $6\%$ below the reported extrapolated values of $6.53$, $7.41$, and $8.04$, respectively. 
We also find a candidate for the $\mathbb{Z}_2$-odd, particle-hole even, $\ell=3$ primary $\sigma'_{\mu_1\mu_2\mu_3}$, with reconstructed dimension $7.39$ compared with the reported value of $7.53$, and a possible counterpart to their excited $\ell=5$ primary $\sigma'_{\mu_1\ldots\mu_5}$ near $\Delta_{\rm rec}\simeq8.7$, compared with $8.37$.\footnote{The $\ell=6$ primary of~\cite{Fardelli:2026zas} lies outside the $\ell$ range shown in Fig~\ref{fig:Low-K2-primaries-L14}.}
We further find low-$|K|^2$ states consistent with some of the primary candidates discarded in~\cite{Fardelli:2026zas}.\footnote{In the $\mathbb{Z}_2$-odd, particle-hole even, $\ell=4$ sector, we find two states near $\Delta_{\rm rec}\simeq8.1$ and $8.8$, consistent with the $\varepsilon$-expansion predictions $8.06$ and $8.90$. 
The lower state is also consistent with the discarded candidate shown in Appendix $D$ of~\cite{Fardelli:2026zas}. 
Similarly, we find a $\mathbb{Z}_2$-even, particle-hole odd, $\ell=5$ state near $8.3$, compared with their discarded candidate extrapolating to $8.42$.}

Our reconstructed spectrum in Figure~\ref{fig:Low-K2-primaries-L14} also shows good agreement with the low-spin primary spectrum reported in~\cite{Zhou:2026afs}, with candidate counterparts across the different symmetry sectors.\footnote{For example, we also find a $\mathbb{Z}_2$-odd, $\ell=1$ candidate at $\Delta_{\rm rec}\simeq8.5$, compared with $9.1$ in~\cite{Zhou:2026afs}.}
Of particular interest, in the $\mathbb{Z}_2$-even, particle-hole even, $\ell=2$ sector, we find three higher-energy low-$|K|^2$ states near $\Delta_{\rm rec}\simeq8.23$, $8.35$, and $9$. Ref.~\cite{Zhou:2026afs} reports two primaries with extrapolated dimensions below $10$ in this energy range with these quantum numbers, and notes that an additional $\ell = 2$ primary predicted by the $\varepsilon$-expansion, with a first-order dimension of $8.11$ (See~\cite{Henriksson:2022rnm}), is missing from their spectrum. 
This raises the possibility that one of our candidates corresponds to the missing operator, although larger system sizes would be needed to establish its identification.

These agreements in the low-energy spectrum are encouraging since both the dilatation operator and the conformal generators are reconstructed directly from the same ground state, without any additional input or tuning using known CFT data.
In particular, we find a similar low-energy operator content already at $L=14$, while Refs.~\cite{Fardelli:2026zas,Zhou:2026afs} obtain their continuum estimates by extrapolating results from larger system sizes.

\section{Toward improved CFT search}\label{sec:improved-search}
Finally, we discuss how the reconstructed conformal generators can help with our longer-term goal of a systematic search for 2+1d CFTs. 
So far, we have used these generators to measure how well a given state satisfies conformal symmetry. 
But this means they can 
be used to guide a search for the best approximation to a CFT groundstate in a given finite Hilbert space.  

In~\cite{Camasca:2026awd}, we proposed the 
square root of the variance of $D_{\rec}$,
\be 
E_{\rm VFPE}\(\ket{\psi}\)
\equiv  \sqrt{ \bra{\psi} ( D_{\rec}^\psi)^2  \ket{\psi} 
- \( \bra{\psi} D_{\rec}^\psi  \ket{\psi}\)^2 }
\ee
(which we call the error of the vector fixed-point equation (VFPE)) as a cost function to search for CFT ground states. 
In the CFT limit the variance of $D_\text{rec}$ and the measures of violation of conformality are both zero, but 
at finite size, we observe a clear competition between the two. 
This motivates supplementing the VFPE with a condition that directly measures the violation of the conformal algebra. 
We therefore consider a multi-objective optimization problem, with the two objectives
\be
    \mathcal{L}(|\psi\rangle) =
\(E_{\rm VFPE}(|\psi\rangle)^2,\,
E_{\rm comm}(|\psi\rangle)^2\),
\ee
where 
\be 
E_{\rm comm} ={1\over \Delta_\sigma^{\rec}} \norm{\left([K_z^{\rm rec},P_z^{\rm rec}]/2-D_{\rm rec}\right)
|\sigma_{\rm rec}\rangle } 
\ee
where $| \sigma_{\rm rec} \rangle$ is 
the first excited state
of $D_{\rec}^\psi$,
and $\Delta_{\sigma}^{\rm rec}$
is its reconstructed energy.  
We use the multiple-gradient descent algorithm (MGDA) \cite{desideri2012multiple}, which combines the gradients of the two objectives to find a direction that decreases both whenever possible.
Here, $|\psi \rangle$ denotes the wavefunction at each step of the gradient descent.
Notice that the cost function only depends on this instantaneous state and no further information: At each step of the gradient descent, we are reconstructing from the state the dilatation operator $D_{\rm rec}$ as well as the generators $P_z^{\rm rec}$ and $K_z^{\rm rec}$ from $\ket{\psi}$.  

\begin{figure}[!ht]
    \centering
    \includegraphics[width=\linewidth]{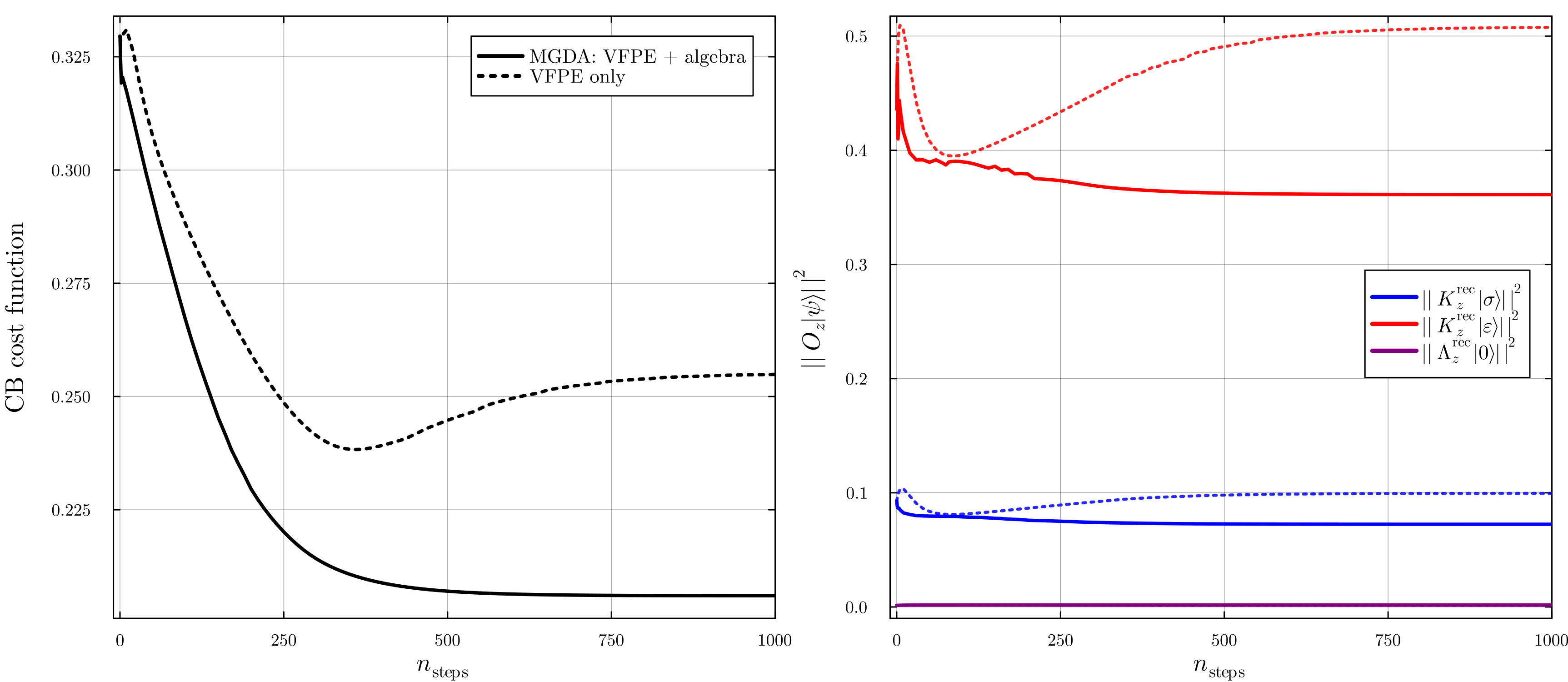}
    \caption{\justifying Improved CFT search for the critical Ising model at $L=6$ with subregions $(A_1,A_2)=(1,2)$. 
    We compare the gradient descent minimizing the error of the VFPE alone (dotted lines) with the MGDA minimizing the combined VFPE and conformal algebra condition (solid lines). 
    Both trajectories start from the same initial state. Left: Cost function defined in~\cite{Camasca:2026awd}, measuring how close the reconstructed spectrum with $\Delta \leq 3$ is to the CB values.
    Right: Squared norms of the reconstructed generators acting on the primaries $\sigma$ and $\varepsilon$, and $\Lambda_z$ acting on the vacuum.}
    \label{fig:gradient-descent-comparison}
\end{figure}

Figure~\ref{fig:gradient-descent-comparison} compares the MGDA descent with minimizing only the error of the VFPE, starting from the same critical Ising ground state at $L=6$.
We see that minimizing only the VFPE initially improves the agreement of the reconstructed spectrum with the conformal bootstrap (CB) values, but the CB cost increases slightly toward the end of the descent procedure.
In contrast, the MGDA descent continues improving the spectrum and stabilizes at a significantly lower CB cost.
We also find that the squared norms of $K_z$ acting on the primaries $\sigma$ and $\varepsilon$ are smaller for the MGDA descent, while for the error of the VFPE descent these quantities initially decrease but eventually increase.
Notice that the improvement extends to $\varepsilon$, even though the commutator condition is imposed only on $\sigma$.
The norm of $\Lambda_z$ acting on the vacuum remains small in both cases.

These results suggest that the conformal algebra provides complementary constraints that can guide the optimization toward more conformal states, even at finite system size.
This offers a promising new ingredient for our quest to find new CFTs in $2+1$ dimensions.  More generally, the reconstruction of the conformal algebra from the groundstate opens slightly wider the window between the maya of what we can study concretely and the real world of CFT.  

\begin{acknowledgments}

Some of our numerics used the FuzzifiED package \cite{zhou2025fuzz}.  We used Julia. 
AI was used for improving our code.
This work was supported in part by
funds provided by the U.S. Department of Energy
(D.O.E.) under cooperative research agreement
DE-SC0009919,
and by the Simons Collaboration on Ultra-Quantum Matter, which is a grant from the Simons Foundation (652264, JM).

\end{acknowledgments}

\clearpage
\let\addcontentsline\savedaddcontentsline %

\onecolumngrid %
\begin{center}
    \textbf{\large Table of Contents for Appendices}
\end{center}
\makeatletter
\print@toc{toc} %
\makeatother
\vspace{2.5em}
\twocolumngrid %

\appendix
\renewcommand{\theequation}
{\Alph{section}.\arabic{equation}}

\section{Review of CFT basics}
\label{app:CFT-basics}

Let us first work in flat $d+1$-dimensional Euclidean space $\mathbb{R}^{d+1}$, with metric 
$g_{\mu\nu}=\delta_{\mu\nu}$.
Conformal transformations are the set of operations on the coordinates $x \to x'(x)$ that satisfy 
$g'_{\mu\nu}(x') = \omega(x)^2 g_{\mu\nu}(x)$.
The infinitesimal generators of such group can be obtained by considering the infinitesimal coordinate transformation $x^\mu \to x^\mu + \epsilon^\mu$, under which the line element transform as:
\begin{equation}
    ds^2 \to ds^2 + (\partial_\mu \epsilon_\nu + \partial_\nu \epsilon_\mu) dx^\mu dx^\nu.
\end{equation}
Since we require this to be a conformal transformation, it implies that:
\begin{equation}
    \partial_\mu \epsilon_\nu + \partial_\nu \epsilon_\mu = \frac{2}{d+1} (\partial\cdot \epsilon) \delta_{\mu \nu}.
\end{equation}
Solutions to this equation are called conformal Killing vectors $v_\epsilon = \epsilon^\mu(x) \partial_\mu$.
Four types of transformations respect this condition for $d+1>2$: translations $\epsilon^\mu = a^\mu$, rotations $\epsilon^\mu = \omega^\mu_\nu x^\nu$, dilations $\epsilon^\mu = \lambda x^\mu$ and special conformal transformations $\epsilon^\mu = 2x^\mu (b\cdot x)-b^\mu x^2$. 

At the CFT level, each of such transformations defines a conserved Noether current,
\begin{equation}
    j^\mu_\epsilon(x) = \epsilon_\nu(x) T^{\mu \nu}(x),
\end{equation}
where $T^{\mu \nu}$ is the energy-momentum tensor. 
The corresponding conserved charge can be evaluated on a
codimension-one surface $\Sigma$,
\begin{equation}
    Q_\epsilon(\Sigma)
    =
    \int_{\Sigma} d\Sigma\,
    n_\mu j^\mu_{\epsilon}
    =
    \int_{\Sigma} d\Sigma\,
    n_\mu\epsilon_\nu T^{\mu\nu}.
\end{equation}
Because $j^\mu_\epsilon$ is conserved, the integration surface can be continuously deformed without changing the charge. 
On a constant-$x^0$ slice in flat Euclidean space this reduces to
\begin{equation}
    Q_\epsilon
    =
    \int d^{d}x\,
    \epsilon_\nu(x)T^{0\nu}(x).
\end{equation}
The conserved charges associated with conformal symmetry are the translation generators $P_\mu$, the rotation generators $M_{\mu\nu}$, the dilatation generator $D$, and the special conformal generators $K_\mu$. 
In Lorentzian signature, $M_{\mu\nu}$ generate Lorentz transformations, whereas in Euclidean signature they generate the $SO(d+1)$ rotation group.

Throughout this work we use the standard Euclidean radial-quantization
convention for the conformal generators.\footnote{Another common convention defines Hermitian generators with additional factors of $i$, in which case, for example, $[D,P_\mu]=iP_\mu$ and $[D,K_\mu]=-iK_\mu$. We instead use the Euclidean radial-quantization convention above, which is convenient because $D^\dagger=D$ and
$P_\mu^\dagger=K_\mu$.}
In this convention, the
non-vanishing commutators of the conformal algebra are:
\begin{align}
    [M^\mu_{\nu},P_\rho] &= \delta_{\nu\rho}P^\mu-\delta^\mu_{\rho}P_\nu, \,\, \\
    [M^\mu_{\nu},K_\rho] &= \delta_{\nu\rho}K^\mu-\delta^\mu_{\rho}K_\nu, \\
    [M^\mu_{\nu},M^\rho_{\sigma}] &= \delta^\rho_{\nu}M^\mu_{\sigma} -\delta^{\mu\rho}M_{\nu\sigma} -\delta_{\nu\sigma}M^{\mu\rho} -\delta^{\mu}_{\sigma}M^\rho_{\nu}. 
\end{align}
Of particular importance for our purposes are the following commutators:
\begin{align}
    [D,P_\mu] &= P_\mu, \label{eq:commutator-D-P}\\
    [D,K_\mu] &= -K_\mu, \label{eq:commutator-D-K} \\
    [K_\mu,P_\nu] &= 2\delta_{\mu\nu}D-2M_{\mu\nu}. \label{eq:commutator-K-P}
\end{align}

We can ask how the field operators $\mathcal{O}_\alpha(x)$ transform under conformal transformations. From the action of $D$ and $M_{\mu \nu}$ at the origin 
\begin{align}
    [D, \mathcal{O}_\alpha(0)] &= \Delta \mathcal{O}_\alpha(0) \\
    [M_{\mu \nu}, \mathcal{O}_\alpha(0)] &= (S_{\mu \nu})_{\alpha}^{\beta} \mathcal{O}_\beta(0)
\end{align}
we see that operators in the CFT are labeled by their scaling dimension $\Delta$ and their spin $\ell$. 
Physically, the scaling dimension corresponds to the exponent in the power law correlation function of the operator $\langle \mathcal{O}(x) \mathcal{O}(0)\rangle \sim 1/ |x|^{2\Delta}$. 
The remaining generators, $P_\mu$ and $K_\mu$, act as raising and lowering operators with increased/decreased scaling dimension by one. We can distinguish between two types of operators: primaries (the highest-weight states, which are annihilated by the lowering operators), and descendants (generated by acting with powers of $P_\mu$ on primaries).

In CFT, a special role is played by the spatial manifold $S^{d}$, because $\mathbb{R} \times S^{d}$ is conformal to $\mathbb{R}^{d+1}$.
The dilatation operator $D$ on $\mathbb{R}^{d+1}$ is proportional to the Hamiltonian on the sphere.
Basis states on $S^{d}$ can be labelled by quantum numbers $(\Delta, \ell)$ 
of the dilatation operator $D$ and the angular momentum, and 
are in one-to-one correspondence with local scaling operators of the CFT. That is, the eigenstates $\ket{\psi_n}$ of the CFT Hamiltonian on $S^{d}$ correspond with local scaling operators, and the energy gaps $\delta E_n$ of these states are proportional to the scaling dimensions $\Delta_n$ of CFT operators $\delta E_n = E_n - E_0 = \frac{v}{R} \Delta_n$, where $R$ is the radius of sphere $S^{d}$ and $v$ is the model-dependent velocity of light. 

\section{Constructing the conformal algebra generators}
\label{app:conformal-generators}

As we mentioned in the introduction, we are interested in understanding 2+1-D CFTs. 
The Hilbert space of a CFT is a representation of
the conformal algebra, generated by the translations $P_\mu$, rotations $M_{\mu \nu}$, dilations $D$, and the special conformal generators $K_\mu$. 
The CFT Hilbert space decomposes into irreducible representations of the conformal algebra, 
which are labelled by $ (\Delta, \ell)$.
Each conformal multiplet is generated from a primary state, which is annihilated by all $K_\mu$, while its descendants are obtained by repeated action of the translation generators $P_\mu$.
Thus, it is important to investigate the conformal algebra on the fuzzy sphere. 
Here we review how the conformal generators can be constructed from the energy-momentum tensor $T_{\mu\nu}$. In particular, we show that the combination $P_i+K_i$ depends only on the energy density $T_{00}$.\footnote{The rotation generators will not be discussed further here. 
Unlike translations and special conformal transformations, the $\gSO(3)$ rotations are exact symmetries of the fuzzy-sphere regularization, and their generators are available from the microscopic angular momentum operators. 
Our main interest will therefore be the generators $P_i$ and $K_i$, together with the dilatation operator $D$.}

Since our interest lies in the conformal generators on the cylindrical geometry $\mathbb{R}\times S^{d}$, we now transform the conserved charges from Euclidean flat space to the cylinder. 
To distinguish the original Cartesian directions from intrinsic coordinates on the sphere, we denote Cartesian indices in $\mathbb{R}^{d+1}$ by $i,j=\{1,\ldots,d+1\}$, while $a,b=\{1,\ldots,d\}$ the intrinsic coordinates on $S^{d}$.
In $d+1$-dimensional Euclidean space, the conformal Killing vectors corresponding to dilatations, translations, and special conformal transformations are
\begin{equation}
    \hspace{-0.1cm}v_D=x^i\partial_i,\,\,
    v_{P_i}=\partial_i,\,\,
    v_{K_i}=-x^2\partial_i+2x_ix^j\partial_j .
\end{equation}
Consider the change of coordinates $x_i = r_{\,}\hat n_i$, where $r = e^\tau = \sqrt{x_i x^i}$ and $\hat n_i \hat n^i = 1$. 
In these coordinates, the flat Euclidean metric becomes
\begin{equation}
    ds^2 = dr^2+r^2d\Omega_{d}^2 = e^{2\tau} \left( d\tau^2+d\Omega_{d}^2 \right).
\end{equation}
Thus, up to the overall Weyl factor, it is the Euclidean cylinder $\mathbb R_\tau\times S^{d}$.
Using the chain rule, the Cartesian derivative becomes: 
\begin{equation}
    \partial_i = e^{-\tau} \left( \hat n_i \partial_\tau + (\delta_i^{j} - \hat n_i \hat n^j) \partial_{\hat n_j} \right).
\end{equation}
And thus, the conformal Killing vectors in the new coordinate system are for dilatations $v_D = \partial_\tau$, and for translations and special conformal transformations: 
\begin{equation}
    \hspace{-0.1cm}v_{P_i} \hspace{-0.15cm}= e^{-\tau} \big( \hat n_i \partial_\tau + \nabla^{(S)}_i \big), \,\, v_{K_i} \hspace{-0.15cm}= e^{\tau} \big( \hat n_i \partial_\tau - \nabla^{(S)}_i \big),
\end{equation}
where 
\begin{equation}
    \nabla^{(S)}_i \equiv (\delta_i^j - \hat n_i \hat n^j) \partial_{\hat n^j}
\end{equation}
projects onto the tangent plane of the sphere.
This same tangential vector field can be written in intrinsic coordinates on $S^{d}$ as
\begin{equation}
    \nabla_i^{(S)} = (\nabla_i^{(S)})^a\partial_a, \qquad (\nabla_i^{(S)})^a = g^{ab}\partial_b\hat n_i,
\end{equation}
where $g_{ab}$ is the metric on the unit sphere.

Using these conformal Killing vectors, we obtain the conserved charges on constant-$\tau$ slices. We evaluate them at $\tau=0$ and Wick rotate to Lorentzian time, under which the energy-current components acquire a factor of $i$. We denote the resulting energy density and currents by $T_{00}$ and $T_{0a}$, respectively.
The dilatation operator is:
\begin{equation}
    D = R \int_{S^{d}} \hspace{-0.3cm} d\Omega\, T_{00} (\Omega), \label{eq:dilation-appendix}
\end{equation}
the conformal generators of translation $P_i$ are: 
\begin{equation}
    P_i = R \int_{S^{d}} \hspace{-0.3cm} d\Omega \left( \hat n_i T_{00}(\Omega) + i(\nabla_i^{(S)})^a T_{0a} \right),
\end{equation}
and the special-conformal generators $K_i$ are:
\begin{equation}
    K_i = R \int_{S^{d}} \hspace{-0.3cm} 
    d\Omega  \left( \hat n_i T_{00}(\Omega) - i (\nabla_i^{(S)})^a T_{0a} \right).
\end{equation}
Notice that these generators satisfy the radial-quantization relation $K_i = P_i^\dagger$.

In practice, it will be useful to define the symmetric combination: 
\begin{equation}
    \Lambda_i \equiv %
    (P_i +K_i) = 2 R \int_{S^{d}} \hspace{-0.3cm} d\Omega \, \hat n_i T_{00}(\Omega). \label{eq:symmetric-generators}
\end{equation}
This is because we can use $\Lambda_i$ and the commutators in~\eqref{eq:commutator-D-P} and~\eqref{eq:commutator-D-K} to extract both $P_i$ and $K_i$ via:
\begin{equation}
    P_i=\frac{1}{2}(\Lambda_i+[D,\Lambda_i]), \qquad K_i=\frac{1}{2}(\Lambda_i-[D,\Lambda_i]). \label{eq:generators-K-P}
\end{equation}
Therefore, once the CFT energy density $T_{00}$ is known, we can construct the dilatation, translation, and special conformal generators using Eqs.~\eqref{eq:dilation-appendix},~\eqref{eq:symmetric-generators}, and~\eqref{eq:generators-K-P}.

The relations above complete the construction of the conformal generators assuming we have direct access to the infrared CFT energy density $T_{00}(\Omega)$. On the fuzzy sphere, however, $T_{00}$ is not known microscopically. Existing approaches approximate the required moments of $T_{00}$ using the microscopic Hamiltonian density~\cite{Fan:2024vcz,Fardelli:2024qla,Fardelli:2026zas,Zhou:2026afs}. Here we take a different approach: we reconstruct these moments directly from the ground state wavefunction, without using the microscopic Hamiltonian density or additional tuning to known CFT data.

\section{Conformal generators from the ground state wavefunction}
\label{app:EB-conformal-generators}

We now construct the global conformal generators on the fuzzy sphere directly from the ground state using an entanglement bootstrap approach. 
We first show that an $\gSO(3)$-weighted average of the modular Hamiltonians isolates the $\ell=1$ moments of the CFT energy density $T_{00}$. 
We then use our model for the UV corrections to show how linear combinations of regions with different sizes can cancel the leading boundary contribution and systematically improve the reconstruction.

We start by considering the modular Hamiltonian of a CFT ground state $|\psi \rangle$ over a spherical region
$A_{\Omega}$ centered at $\Omega$, with opening angle $\theta_A$:
\begin{equation}
    K^{\mathrm{CFT}}_{A_{\Omega}} =\int_{A_\Omega} d\Omega_x\,\beta_A\!\left(\gamma(\Omega,x)\right) T_{00}(x) +f_A\mathbb{I},
\label{eq:KA-CFT}
\end{equation}
where $\gamma(\Omega,x)$ is the angle between the center $\Omega$ and the
point $x$ on the sphere. The coolness function is:
\begin{equation}
    \beta_A(\gamma) = 4\pi R \, \frac{\sin(\frac{1}{2}(\theta_A-\gamma)) \sin(\frac{1}{2}(\theta_A+\gamma))}{\sin\theta_A}\,\Theta(\theta_A-\gamma),
\label{eq:beta-half-angle}
\end{equation}
with $f_A=\frac{|\partial A|}{\epsilon}-F$, and the boundary of the spherical region $|\partial A|=2\pi R\sin\theta_A$.

On a regulated system, the modular Hamiltonian (actually exists and) also contains extra UV terms. 
We expect that the entanglement Hamiltonian is of the form $K_{A_\Omega} = K_{A_\Omega}^{CFT} + K_{A_\Omega}^{UV}$. 
Since the UV term captures the short-distance entanglement, its contribution is localized at the boundary $\partial A_{\Omega}$:
\begin{equation}
    K^{\mathrm{UV}}_{A_{\Omega}} =\int_{\partial A_{\Omega}}d\ell\, \left( O_0^{\hat n}(x) +\kappa \, O_1^{\hat n}(x)+\cdots \right), \label{eq:KUV-local}
\end{equation}
where $\hat n$ is the outward normal to the boundary within the sphere, $\kappa = \nabla_a n^a$ is the boundary curvature, and $O^{\hat n}_i(x)$ are local boundary operators. 
For a spherical cap of opening angle $\theta_A$, $\kappa = \cot \theta_A /R$.

The question is therefore: Given our UV model for the modular Hamiltonian on the fuzzy sphere, can we figure out a way as to extract $\Lambda_i$ defined on~\eqref{eq:symmetric-generators} and thus construct the conformal generators? 
Note that a similar question was posed in~\cite{Camasca:2026awd}, where extracting the CFT dilatation operator $D^{\mathrm{CFT}} = R \int_{S^2} d\Omega \, T_{00}(\Omega)$ through $\gSO(3)$ averaging and a linear combination of $k=2$ regions yielded successful results.
We will see below that the answer to such question is a parallel procedure where we do a weighted $\gSO(3)$ averaging to isolate the $\Lambda_i$ term and take a linear combination of $k=2$ regions to cancel the leading UV term.  

Let us first show how we can extract $\Lambda_i$ from the CFT ground state. If $| \psi\rangle$ is the ideal CFT ground state, where the modular Hamiltonian takes the form~\eqref{eq:KA-CFT}, we could average it over the center $\Omega$ against a spherical-harmonic $\overline K ^{\mathrm{CFT}}_{A,(\ell,m)} \equiv \int d\hat\Omega\,Y_{\ell m}^{*}(\Omega) K^{\mathrm{CFT}}_{A_{\Omega}} $: 
\begin{align}
    \hspace{-0.2cm}\overline K ^{\mathrm{CFT}}_{A,(\ell,m)} = \int_{S^2} \hspace{-0.2cm}d\Omega_x \hspace{-0.05cm} \int \hspace{-0.05cm}d\hat\Omega\, Y_{lm}^*(\Omega) \beta(\gamma) T_{00}(x) + \frac{f_A \delta_{l,0}}{\sqrt{4\pi}}.\label{eq:weighted-average-def}
\end{align}
Here $d\Omega_x$ denotes the angular measure on $S^2$, while $d\widehat{\Omega}\equiv d\Omega/(4\pi)$ is the normalized measure over the cap centers, with $\int_{S^2}d\widehat{\Omega}=1$. The functions $Y_{\ell m}(\Omega)$ are the spherical harmonics of spin $\ell$ and magnetic quantum number $m$.\footnote{Up to a prefactor, $l=0$ averaging gives $D^{\mathrm{CFT}} = R \int d\Omega T_{00}(\Omega)$.}\textsuperscript{,}\footnote{In $D=2+1$, the $l=\{0,1\}$ components are special, as they are related to the global conformal generators and hence annihilate the CFT ground state. 
For $l\geq 2$, there is no generator of the finite-dimensional conformal algebra with the corresponding angular quantum numbers, and hence no reason a priori for these weighted operators to annihilate the vacuum. 
This differs from $D=1+1$ dimensions, where local conformal symmetry is infinite dimensional and higher Fourier modes can be organized in terms of Virasoro generators.
We leave the study of larger-$\ell$ components for future work.}
To evaluate this weighted average, we expand the coolness function in Legendre polynomials,
\begin{equation}
    \beta_A(\gamma) =\sum_{l=0}^{\infty}\frac{2l+1}{2}\,a_l P_l(\cos\gamma), \label{eq:beta-Legendre}
\end{equation}
with
\begin{equation}
    a_l = \int_0^\pi d\gamma\,\sin\gamma\, P_l(\cos\gamma)\beta_A(\gamma).
\end{equation}
Using the addition theorem for spherical harmonics $P_l(\cos \gamma) =\frac{4\pi}{2l+1}
\sum_{m=-l}^{l}Y_{lm}(\Omega)Y_{lm}^{*}(x)$, 
and the orthogonality of the spherical harmonics, we find that
\begin{equation}
\int d\hat\Omega\,Y_{\ell m}^{*}(\Omega)
\beta_A\!\left(\gamma(\Omega,x)\right)
=\frac{a_\ell}{2}Y_{\ell m}^{*}(x).
\label{eq:weighted-kernel}
\end{equation}
Substituting this result into Eq.~\eqref{eq:weighted-average-def}, we obtain
\begin{equation}
\overline K^{\mathrm{CFT}}_{A,(\ell,m)}
=R v_\ell(\theta_A)
\int_{S^2} d\Omega_x\,Y_{\ell m}^{*}(x)T_{00}(x)
+\frac{f_A}{\sqrt{4\pi}}\delta_{l,0},
\label{eq:weighted-CFT-final}
\end{equation}
where
\begin{equation}
v_\ell(\theta_A)
\equiv\frac{a_\ell}{2R}
=\frac{1}{2R}\int_0^\pi d\gamma\,\sin\gamma\,
P_\ell(\cos\gamma)\beta_A(\gamma).
\label{eq:v-l-def}
\end{equation}
To obtain the conformal generators, we set $\ell=1$. Since $P_1(\cos\gamma)=\cos\gamma$, the coefficient is given by:
\begin{equation}
v_1(\theta_A) = 2\pi  \left( \frac{\sin^4(\theta_A/2)}{\sin\theta_A} \right)\left(\frac{2+\cos\theta_A}{3}\right).
\label{eq:v1}
\end{equation}
Hence,
\begin{equation}
\overline K^{\mathrm{CFT}}_{A,(1,m)}
=v_1(\theta_A)
R \int_{S^2} d\Omega_x\,Y_{1m}^{*}(x)T_{00}(x).
\label{eq:l1-CFT}
\end{equation}
The three $\ell=1$ moments determine the spherical components of
$\Lambda_i$.

In the CFT limit, Eq.~\eqref{eq:l1-CFT} reconstructs the conformal generators. 
However, in the regulated system, the modular Hamiltonian also contains UV contributions localized at the entangling surface. 
We will show that $k=2$ different region sizes are enough to cancel the leading UV contribution. 
To see this, we apply the same weighted average to the UV contribution $\overline K ^{\mathrm{UV}}_{A,(\ell,m)} \equiv \int d\hat\Omega\,Y_{\ell m}^{*}(\Omega) K^{\mathrm{UV}}_{A_{\Omega}}$, which gives:
\begin{equation}
    \overline K^{\mathrm{UV}}_{A,(\ell,m)} = \int d\hat\Omega\, Y_{l m}^{*} \!\left( \Omega\right) \int_{\partial A_\Omega}\hspace{-0.3cm}d\ell \left(O_0^{\hat n}(x) +\kappa \,O_1^{\hat n}(x)+\cdots\right).
\end{equation}
We can equivalently parametrize this integral by first choosing a point $x$ on the sphere and then a unit tangent vector $\hat n$ corresponding to the outward normal of a cap $A_\Omega$ whose boundary passes through $x$. Since $x\in\partial A_\Omega$, the angle between the cap center $\Omega$ and $x$ is $\theta_A$, so that $\Omega\cdot x=\cos\theta_A$. Decomposing $\Omega$ into its component along $x$ and its component tangent to the sphere at $x$, and noting that the latter points opposite to the outward normal $\hat n$, we obtain
\begin{equation}\label{eq:thetaA-and-n}
    \Omega=\cos\theta_A\, x-\sin\theta_A\,\hat n.
\end{equation}
Thus, fixing $(x,\hat n)$ uniquely determines the center of the corresponding cap. Moreover, since the boundary circle has radius $R\sin\theta_A$, we have $d\ell=R\sin\theta_A\,d\phi$. 
We can therefore exchange the integration over cap centers and their boundaries for an integration over all points $x$ and tangent directions $\hat n$: 
\begin{align}
    \overline K^{\mathrm{UV}}_{A,(\ell,m)} =R\sin\theta_A \int_{S^2} d\Omega_x \int d\hat n\, Y_{l m}^{*} \!\left(\cos\theta_A x- \sin\theta_A\hat n\right) \times \nonumber\\ 
    \left(O_0^{\hat n}(x) +\kappa \,O_1^{\hat n}(x)+\cdots\right).
\label{eq:weighted-UV-general}
\end{align}
Here $\int d\hat n\equiv 1/(4\pi)\int_0^{2\pi}d\varphi_{\hat n}$ denotes the integration over tangent-normal directions at fixed $x$, including the normalization of the average over cap centers.

The formula in Eq.~\eqref{eq:weighted-UV-general} is true for general $(l,m)$. 
We are interested in the $l=1$ component. In this case, the spherical harmonics are linear and satisfy: 
\begin{equation}
Y_{1m}^{*}
\!\left(\cos\theta_A x-
\sin\theta_A\hat n\right)
=\cos\theta_A\,Y_{1m}^{*}( x)
-\sin\theta_A\,Y_{1m}^{*}(\hat n).
\label{eq:Y1-linear}
\end{equation}
Substituting this into Eq.~\eqref{eq:weighted-UV-general} gives
\begin{align}
\overline K^{\mathrm{UV}}_{A,(1,m)}
=&R\sin\theta_A\cos\theta_A
\int_{S^2} d\Omega_x
\int d\hat n\,
Y_{1m}^{*}( x)O_0^{\hat n}(x) \nonumber \\
&-R\sin^2\theta_A
\int_{S^2} d\Omega_x
\int d\hat n\,
Y_{1m}^{*}(\hat n)O_0^{\hat n}(x)
\nonumber\\
&+\kappa R\sin\theta_A\cos\theta_A
\int_{S^2} d\Omega_x \int d\hat n\, Y_{1m}^{*}( x)O_1^{\hat n}(x) \nonumber \\
&-\kappa R\sin^2\theta_A \int_{S^2} d\Omega_x \int d\hat n\, Y_{1m}^{*}(\hat n)O_1^{\hat n}(x) +\mathcal{O}(\kappa^2R).
\label{eq:weighted-UV-expanded}
\end{align}
Using $O_i^{\hat n}(x)=(-1)^iO_i^{-\hat n}(x)$ and $Y_{1m}(-\hat n)=-Y_{1m}(\hat n)$, we see that:
\begin{equation}
\int d\hat n\,Y_{1m}^{*}(\hat n)O_0^{\hat n}(x) = 
\int d\hat n\,Y_{1m}^{*}( x)O_1^{\hat n}(x)=0.
\label{eq:vanishing-weighted}
\end{equation}
This leaves us with:
\begin{align}
    \overline K^{\mathrm{UV}}_{A,(1,m)} &=R\sin\theta_A\cos\theta_A\,\int_{S^2} d\Omega_x \int d\hat n\,Y_{1m}^{*}( x)O_0^{\hat n}(x) \nonumber \\ &
    - \kappa R\sin^2\theta_A\,\int_{S^2} d\Omega_x \int d\hat n\,Y_{1m}^{*}(\hat n)O_1^{\hat n}(x) +\mathcal{O}(\kappa^2R),
\label{eq:weighted-UV-surviving}
\end{align}
For fixed opening angles, the leading UV terms in Eq.~\eqref{eq:weighted-UV-surviving} scale as $O(R)$ and $O(1)$, respectively. 
However, both have the same dependence on the subregion size, proportional to $\sin\theta_A\cos\theta_A$. 
Thus, the first condition in Eq.~\eqref{eq:Lambda-k2-conditions}, 
cancels both contributions to the weighted average defined in Eq.~\eqref{eq:cartesian-weighted-modular-Hamiltonian}, while the second condition fixes the normalization of the CFT generator. 

We can further improve the reconstruction by canceling
the next UV contribution in the curvature expansion.
To determine its dependence on the region size, let us consider the next subleading UV terms from the curvature expansion in Eq.~\eqref{eq:KUV-local}:
\begin{align}
    \Lambda_i^{\rm rec} ={}& \Lambda_i
    + \sum_\alpha 2\tilde\lambda_{A_\alpha}\sin\theta_{A_\alpha}\cos\theta_{A_\alpha}
      \sum_{\ell\ge0}\kappa_{A_\alpha}^\ell \nonumber\\
    &\qquad\times\int R\,d\Omega_x\int d\hat n\, x_i\, O_\ell^{\hat n}(x) \nonumber\\
    &- \sum_\alpha 2\tilde\lambda_{A_\alpha}\sin^2\theta_{A_\alpha}
      \sum_{\ell\ge1}\kappa_{A_\alpha}^\ell \nonumber\\
    &\qquad\times\int R\,d\Omega_x\int d\hat n\, \hat n_i\, O_\ell^{\hat n}(x) \nonumber\\
    ={}& \Lambda_i
    + \sum_{\ell\ge0}\frac{2\tilde G_\ell(\{A_\alpha\})}{R^{\ell-1}}
      \int d\Omega_x\int d\hat n\, x_i\, O_\ell^{\hat n}(x) \nonumber\\
    &- \sum_{\ell\ge1}\frac{2\tilde F_\ell(\{A_\alpha\})}{R^{\ell-1}}
      \int d\Omega_x\int d\hat n\, \hat n_i\, O_\ell^{\hat n}(x),
    \label{eq:Lambda-rec-UV-expansion}
\end{align}
where in the second equality we used $\kappa_{A_\alpha}=\cot\theta_{A_\alpha}/R$ for the extrinsic curvature of the boundary of the cap $A_\alpha$.
The subregion dependence is contained in:
\be
    \tilde G_\ell(\{A_\alpha\}) = \sum_\alpha \tilde \lambda_{A_\alpha} \sin \theta_{A_\alpha} \cos \theta_{A_\alpha} \cot^\ell \theta_{A_\alpha},
\ee
and 
\be
    \tilde F_\ell(\{A_\alpha\}) = \sum_\alpha \tilde \lambda_{A_\alpha} \sin^2 \theta_{A_\alpha} \cot^\ell \theta_{A_\alpha}.
\ee

Using the parity property in Eq.~\eqref{eq:parity-relation-for-CO}, only
even $\ell$ contribute to the terms proportional to
$\tilde G_\ell$, while only odd $\ell$ contribute to those
proportional to $\tilde F_\ell$. 
Moreover, the
coefficients satisfy
\begin{equation}
\tilde F_{\ell+1}(\{A_\alpha\})
=\tilde G_\ell(\{A_\alpha\}).
\end{equation}
Thus, the condition $\widetilde G_0=0$ already cancels
the contribution proportional to $\widetilde F_1$.
The next UV contributions are proportional to
$\widetilde G_2$ and $\widetilde F_3=\widetilde G_2$,
and can both be canceled by imposing
\begin{equation}
\widetilde G_2(\{A_\alpha\})
=\sum_\alpha\tilde\lambda_{A_\alpha}\hspace{-0.05cm}
\sin\theta_{A_\alpha}\hspace{-0.075cm}\cos\theta_{A_\alpha}\hspace{-0.075cm}
\cot^2\theta_{A_\alpha}\hspace{-0.1cm}=0.
\end{equation}
Together with the two conditions in Eq.~\eqref{eq:Lambda-k2-conditions},
this determines the coefficients for the improved
reconstruction with $k=3$ regions.

\begin{figure}[!ht]
    \centering
    \includegraphics[width=\linewidth]{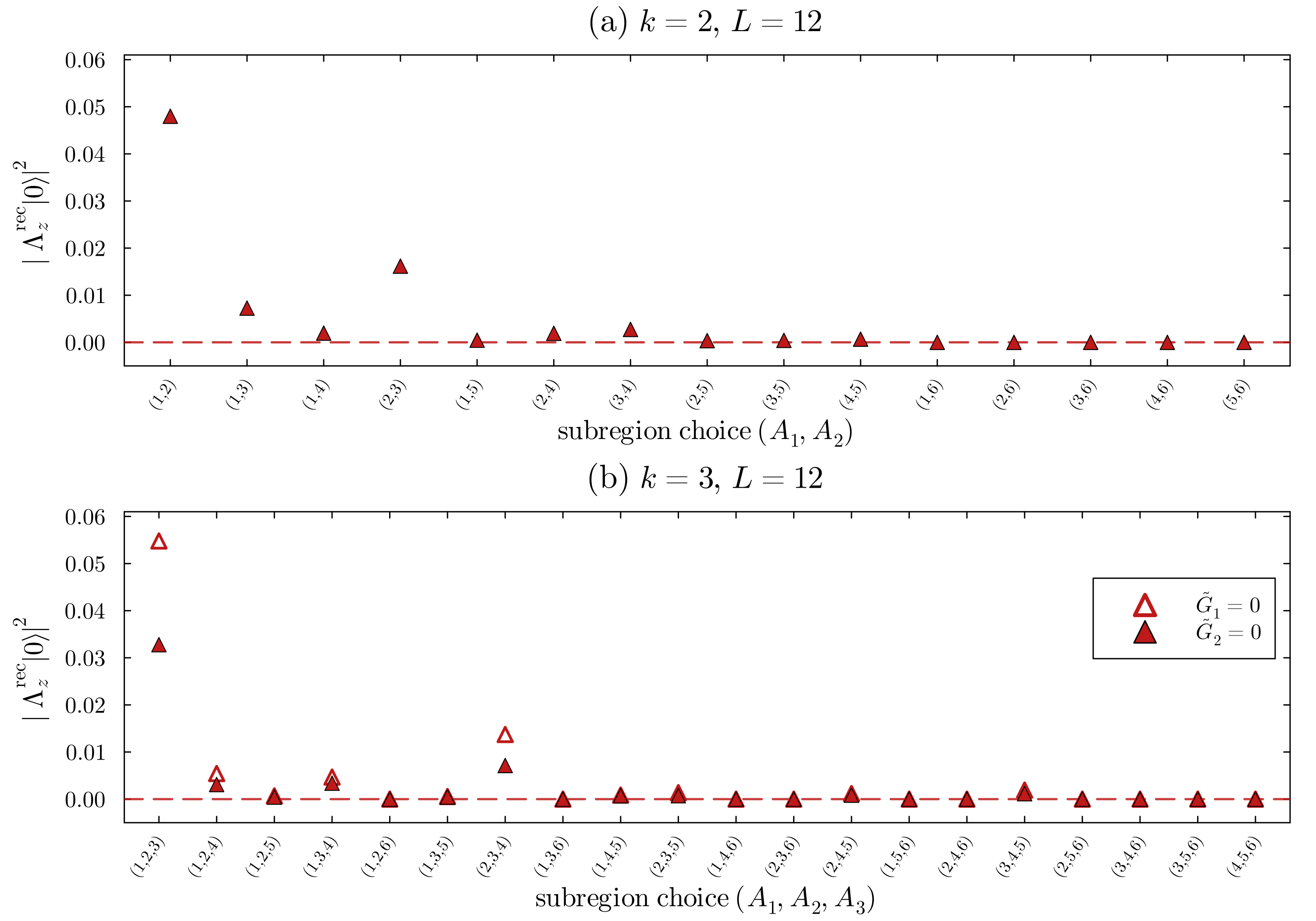}
    \caption{\justifying Squared norm of the reconstructed conformal boost $|| \Lambda_z^{\rm rec} | 0 \rangle||^2$ acting on the critical ground state at $L = 12$ as a function of subregion size. 
    The horizontal dashed line denotes the CFT expectation of zero. 
    $(a)$ Reconstruction using $k=2$ regions. 
    $(b)$ Reconstruction using $k = 3$ regions. Open and filled triangles correspond to the additional condition being $\tilde G_1= 0$ and $\tilde G_2 = 0$, respectively. 
    Imposing $\tilde G_2=0$ cancels the terms proportional to $\tilde G_2/R$ and $\tilde F_3/R^2$
    and visibly improves the results.
}
    \label{fig:Lambda-rec-comparison-k}
\end{figure}

To test this improvement, we examine the action of the reconstructed conformal boost on the critical Ising ground state. 
In the CFT limit, the conformal generators annihilate the vacuum, so $\|\Lambda_z|0\rangle\|^2=0$. 
Figure~\ref{fig:Lambda-rec-comparison-k} shows the dependence of this quantity on the subregion choice at fixed system size $L=12$. 
For $k=3$, we compare imposing $\tilde G_1=0$, which does not cancel the leading remaining UV correction, with the predicted condition $\tilde G_2=0$. 
We find that imposing $\tilde G_2=0$ reduces the deviations from zero for several choices of smaller subregions, while for larger subregions both prescriptions give results close to zero. 
This provides further support for our model of the UV corrections to the modular Hamiltonian.

\section{Numerical implementation}

We follow the numerical procedure described in Appendix A of~\cite{Camasca:2026awd} to construct the modular Hamiltonian $K_A$ and the $\gSO(3)$ irreducible representations of the fuzzy sphere Hilbert space. 
Here we focus on the additional steps needed to compute the weighted averages introduced in Appendix~\ref{app:EB-conformal-generators}. 

{\bf Weighted $\gSO(3)$ averaging.} Let $K_A$ denote the modular Hamiltonian for a spherical region centered at the north pole. The weighted average is defined as:
\begin{equation}
    \overline K_{A, (\ell,m)}=\int_{S^2} d \widehat \Omega \,Y_{\ell m}^*(\Omega)U(\Omega)K_A U(\Omega)^{\dagger},
\label{eq:rotated-KF}
\end{equation}
where $U(\Omega)$ rotates the north pole to $\Omega$.

To evaluate this average numerically, we use the $\gSO(3)$ irreducible representations of the Hilbert space as in \cite{Camasca:2026awd}. 
Therefore, we work in a basis $|s,m_1,\alpha\rangle$, where $s$ is the spin, $m_1$ its $z$-component and $\alpha$ labels multiplicities. Under rotations, these states transform as: 
\be
    U(g)|s,m_1,\alpha\rangle = \sum_{\tilde m}|s,\tilde m,\alpha\rangle D_{\tilde m m_1}^{(s)}(g), \label{eq:U-action}
\ee
where $D^{(s)}(g)$ denotes the Wigner-$D$ matrix. 

Since $K_A$ is centered at the north pole, it is invariant under rotations around the $z$ axis, and its matrix elements are diagonal in $m_1$. Using Eq.~\eqref{eq:U-action}, the matrix elements of the rotated modular Hamiltonian can therefore be written as
\begin{equation}
\begin{aligned}
&\langle s,m_1,\alpha|
U(g)K_AU(g)^\dagger
|s',m_2,\beta\rangle
\\
&=\sum_{\widetilde m=-\min(s,s')}^{\min(s,s')}
D_{m_1\widetilde m}^{(s)}(g)
D_{m_2\widetilde m}^{(s')*}(g)
\langle s,\widetilde m,\alpha|
K_A
|s',\widetilde m,\beta\rangle.
\end{aligned}
\label{eq:rotated-K-matrix}
\end{equation}
This expression separates the matrix elements of the north-pole modular Hamiltonian from the angular dependence, which is entirely contained in the Wigner-$D$ matrices. Thus, using $Y_{\ell m}^{*}(\Omega)=\sqrt{(2\ell+1)/(4\pi)}\,D_{m0}^{(\ell)}(g)$, the weighted average in Eq.~\eqref{eq:rotated-KF} reduces to an integral of three Wigner-$D$ matrices.

To perform the angular integration, we decompose the product of two Wigner-$D$ matrices
\be
    D_{m_1\tilde m}^{(s)}(g)D_{m0}^{(\ell)}(g) =\sum_{J,M,N} C^{J,M}_{s,m_1;\ell,m} C^{J,N}_{s,\tilde m; \ell,0} D_{MN}^{(J)}(g), \label{eq:D-product}
\ee
where $C^{J,M}_{s,m_1;\ell,m} = \langle J,M| s,m_1; \ell,m \rangle$ is the Clebsch-Gordan coefficient.
Using the orthogonality of the Wigner-$D$ matrices, the remaining angular integral selects $J=s'$, giving the matrix elements of the weighted average:
\begin{align}
\langle s,&m_1,\alpha|\overline K_{A,(\ell,m)}|s',m_2,\beta\rangle = \delta_{m_2, m_1+m} \sqrt{\frac{2\ell+1}{4\pi}}\,
\frac{C^{s',m_2}_{s,m_1;\ell,m}}{2s'+1}   \nonumber\\
&  \times\sum_{\tilde m = - \mathrm{min}(s,s')}^{\mathrm{min}(s,s')}
C^{s',\tilde m}_{s,\tilde m;\ell,0}
\langle s,\tilde m,\alpha|K_{A}|s',\tilde m,\beta\rangle.
\label{eq:final-matrix}
\end{align}
The Clebsch-Gordan coefficients enforce the selection rules $m_2 = m_1+m$ and $|s-\ell|\leq s' \leq s+\ell$.
Thus, the weighted average can be computed directly from the matrix elements of the north-pole modular Hamiltonian, without explicitly performing the angular integration numerically.\footnote{Notice that setting $(\ell,m)=(0,0)$ reproduces the ordinary $\gSO(3)$ average used to construct $D_{\rm rec}$, up to a factor of $1/\sqrt{4\pi}$.} 

For the conformal generators, we are interested in $\ell=1$ weighted averages. 
Equation~\eqref{eq:final-matrix} shows that the dependence on $m$ is entirely contained in the Clebsch--Gordan coefficients. Hence, for each pair of $\gSO(3)$ irreducible representations, we can compute the corresponding reduced matrix elements and use them to obtain all three components $m=-1,0,1$. These are then combined for different region sizes to construct $\Lambda_i^{\rm rec}$, as described in Appendix~\ref{app:EB-conformal-generators}.

{\bf Selection rules.}
The numerical calculation can be simplified further by using the symmetries of the ground state to identify which blocks of the weighted modular Hamiltonian can be nonzero. 
We illustrate this for the critical Ising model, extending the analysis of the $\gSO(3)$ average in Appendix A of~\cite{Camasca:2026awd}.
Similar arguments apply to the other critical models and their corresponding symmetries.

Besides spatial $\gSO(3)$ rotations, the Ising model has two discrete symmetries: the Ising $\mathbb Z_2$ and particle-hole symmetry. 
The ground state is a rotational singlet and is invariant under both discrete symmetries. 
We label the $\gSO(3)$ multiplets by $|s,m_1,z_2,p,\alpha\rangle$, where $z_2=\pm1$ and $p=\pm1$ are the corresponding symmetry labels. 
In particular, particle-hole symmetry reverses the magnetic quantum number $m_1$,
\begin{equation}
U_{\rm PH}|s,m_1,z_2,p,\alpha\rangle
=p|s,-m_1,z_2,p,\alpha\rangle.
\label{eq:PH-transformation}
\end{equation}

These symmetries impose additional constraints on the matrix elements in Eq.~\eqref{eq:final-matrix}. 
As discussed in~\cite{Camasca:2026awd}, the Ising $\mathbb Z_2$ symmetry implies that $K_A$ preserves $z_2$, so the matrix elements vanish unless $z_2'=z_2$. 
Particle-hole symmetry, on the other hand, relates the contributions with opposite $\tilde m$. 
Using Eq.~\eqref{eq:PH-transformation} and the Clebsch--Gordan identity
$C^{s',-\tilde m}_{s,-\tilde m;\ell,0}
=(-1)^{s+\ell-s'}C^{s',\tilde m}_{s,\tilde m;\ell,0}$,
we find that the two contributions differ by a phase $pp'(-1)^{s+\ell-s'}$. Hence, the sum over $\tilde m$ vanishes unless
\begin{equation}
pp'(-1)^{s+\ell-s'}=1.
\label{eq:PH-selection}
\end{equation}
For $\ell=0$, this reproduces the selection rule $p'=p$ in~\cite{Camasca:2026awd}. For the $\ell=1$ weighted average, where $s'=s,s\pm1$, we obtain
\begin{equation}
\begin{aligned}
s'=s\pm1 &: \qquad p'=p,\\
s'=s &: \qquad p'=-p.
\end{aligned}
\label{eq:PH-selection-l1}
\end{equation}
The $s'=s$ contribution vanishes for $m=0$ and $m_1=m_2=0$, but is needed to reconstruct all three components of $\Lambda_i^{\rm rec}$.

\section{Results for reconstruction with full modular Hamiltonians}
\label{app:BW}

We now test the alternative construction of the dilatation operator and the conformal generators using full modular Hamiltonians, as discussed in Sec.~\ref{sec:strategies}.
Unlike the averaging method, this construction does not require averaging over the centers of the subregions, which could allow us to study larger system sizes. 
Moreover, the UV contributions localized near the common boundary $\partial A$ of a region $A$ and its complement $\bar A$ are expected to cancel in the full modular Hamiltonian.

\begin{figure}[!t]
    \centering
    \includegraphics[width=\linewidth]{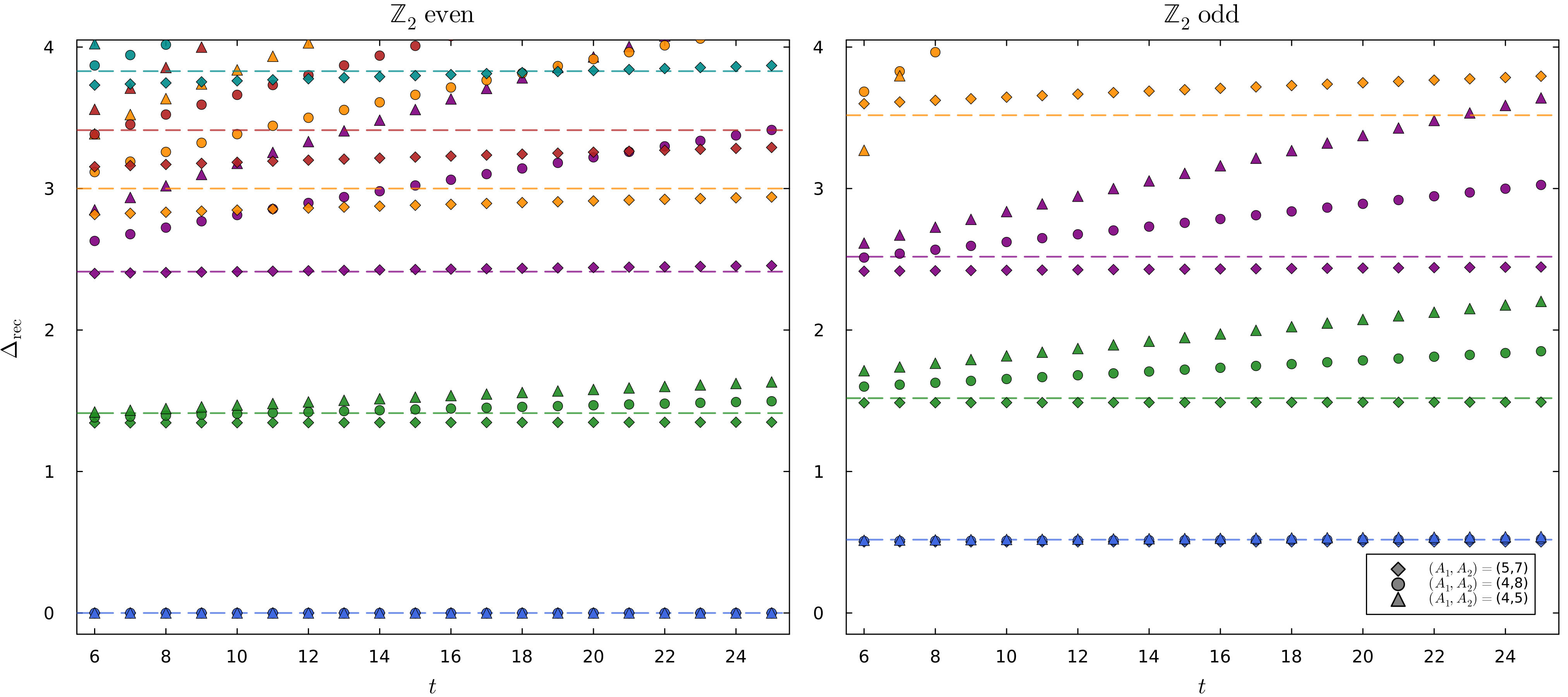}
    \caption{\justifying Reconstructed low-energy spectra of $D_{\rm rec}^{\rm FM}$ obtained from full modular Hamiltonians for the critical Ising model at $L=12$, for different subregion choices. 
    We diagonalize $D_{\rm rec}^{\rm FM}$ within each angular momentum sector.
    We regularize the zero eigenvalues $\lambda < \delta = 10^{-12}$ of the reduced density matrix by assigning a finite value to $-\ln \lambda$, which we vary through the regularization strength $t$.  
    We also checked that changing the cutoff $\delta$ or the numerical precision has a negligible effect on the low-energy spectrum. 
    Horizontal lines correspond to the CB values.
    In each symmetry sector, we show as many states as there are CB levels.
    }
    \label{fig:Drec-FM-Ising}
\end{figure}

We first consider the reconstruction of the dilatation operator using two full modular Hamiltonians with different opening angles.
One difficulty with this construction is that the reduced density matrices can have zero eigenvalues, making the modular Hamiltonian ill-defined. 
This is particularly important for unequal bipartitions, where the larger subsystem has additional zero eigenvalues.
We regularize these following the prescription described in Appendix A of~\cite{Camasca:2026awd}, and diagonalize $D^{\rm FM}_{\rm rec}$ within each angular momentum sector $\ell$.
Figure~\ref{fig:Drec-FM-Ising} shows the reconstructed low-energy spectrum of the critical Ising model at $L=12$ for three subregion choices, as a function of the regularization strength $t$.
We find that the very low-energy levels are stable as we vary $t$, particularly for the closest to the hemisphere $(A_1,A_2)= (5,7)$, while higher energy levels are more sensitive. 
We emphasize that no additional rescaling is used here.

However, this method has the severe drawback that the resulting eigenstates of $D^{\rm FM}_{\rec}$ all have a nonzero overlap with the (effective) kernel of $\rho_A$ where $A$ is some region involved in the construction of $D_\rec$.  
In the effective kernel we include not only eigenvectors of $\rho_A$ with eigenvalue exactly zero, but also eigenvectors with eigenvalue less than the cutoff $\delta$ required by the finite the numerical precision, \ie~to avoid underflow.  
Call the projector onto this effective kernel $P_A$.  
This overlap $ \bra{E} P_A \ket{E}$ varies with $E$ and is quite small for low-energy states.
This has the consequence that the spectrum varies (approximately linearly) with regularization strength $t$ with a slope determined by the overlap (see Figure~\ref{fig:Drec-FM-Ising}).  
This makes it difficult to use this method to make sharp statements about the spectrum.

\begin{figure}[!ht]
    \centering
    \includegraphics[width=0.9\linewidth]{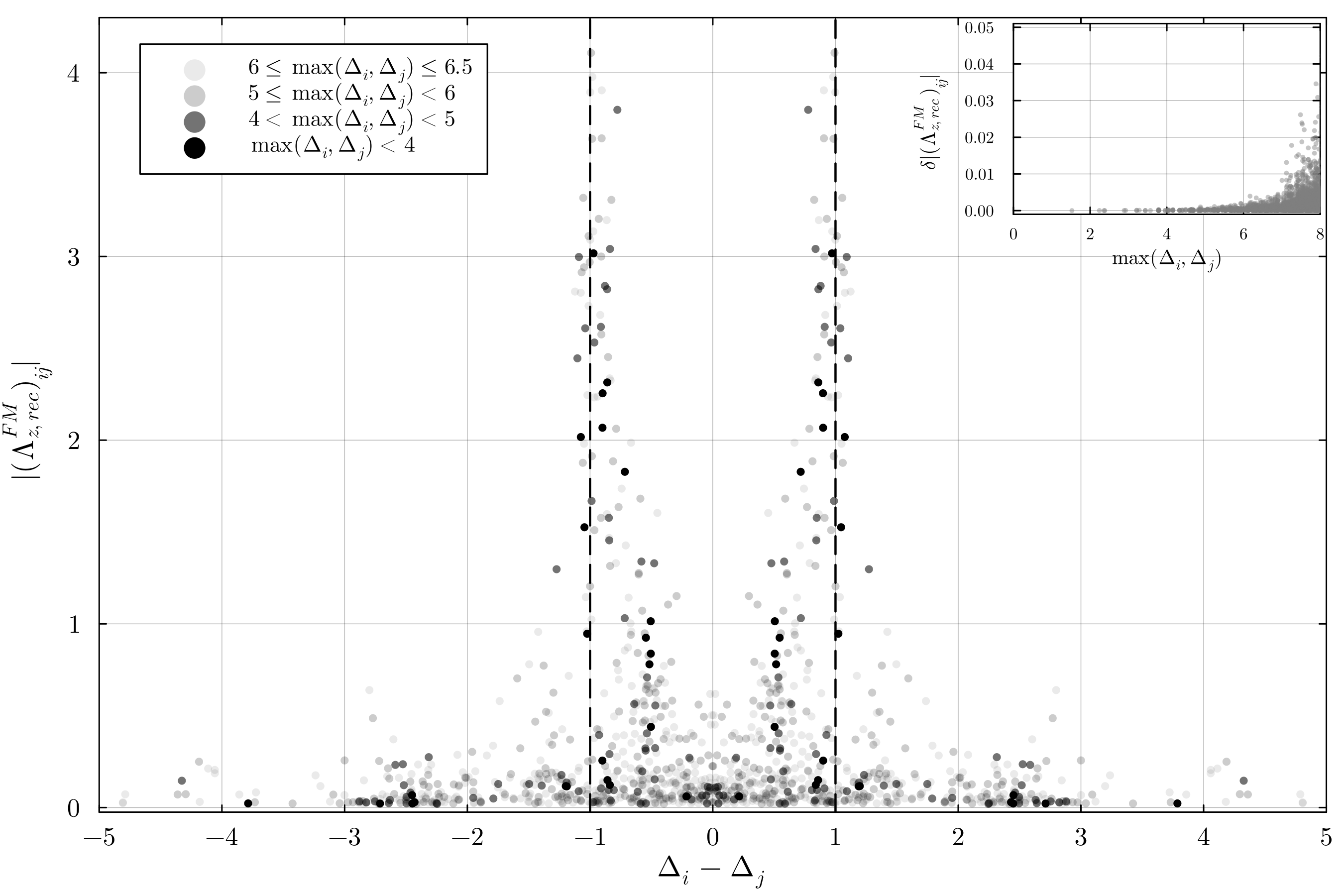}
    \caption{\justifying Matrix elements of the conformal boost reconstructed from the full modular Hamiltonian for the critical Ising model at $L=12$, evaluated between the eigenstates of $D_{\rm rec}$. 
    Inset: Change in the magnitude of the matrix elements between two prescriptions for zero eigenvalues $\lambda$: $\ln \lambda = 0$ or $\ln \lambda = \ln \lambda_{\min}$ where $\lambda_{min}$ is the smallest nonzero eigenvalue of $\rho_A$.} 
    \label{fig:matrix-elements-Ising-proj-BW}
\end{figure}

We next consider reconstructing the conformal boost using a single hemispherical region.
In this case, the full modular Hamiltonian is proportional to $\Lambda_z$ so we do not need to combine different regions.
Figure~\ref{fig:matrix-elements-Ising-proj-BW} shows the matrix elements of the reconstructed $\Lambda_{z,{\rm rec}}^{\rm FM}$ in the eigenbasis of the averaged reconstructed dilatation operator $D_{\rm rec}$. 
The largest matrix elements are concentrated near the expected $\Delta_i - \Delta_j = \pm 1$. 
We also find that these matrix elements depend weakly on the prescription used for the zero eigenvalues, while the differences become larger at higher energies.

\section{Results for other CFTs}
\label{app:other-CFTs}
In this appendix, we test the claim that our method of reconstructing the conformal algebra generators is model independent.  We test our reconstruction on more approximate CFT groundstates: the free scalar, the $O(2)$ Wilson-Fisher, and the $O(3)$ Wilson-Fisher models studied in Refs.~\cite{He:2025ong,Guo:2025odn,dey2026:O3}.
Starting from their critical ground states, we reconstruct both $D_{\rm rec}$ and $\Lambda^{\rm rec}_i$ using the same procedure as we used for the Ising model.
We then examine whether the reconstructed generators approximately satisfy the conformal algebra and whether they can be used to identify primary states in the CFT spectrum. 
\begin{figure}[!ht]
    \centering
    \includegraphics[width=\linewidth]{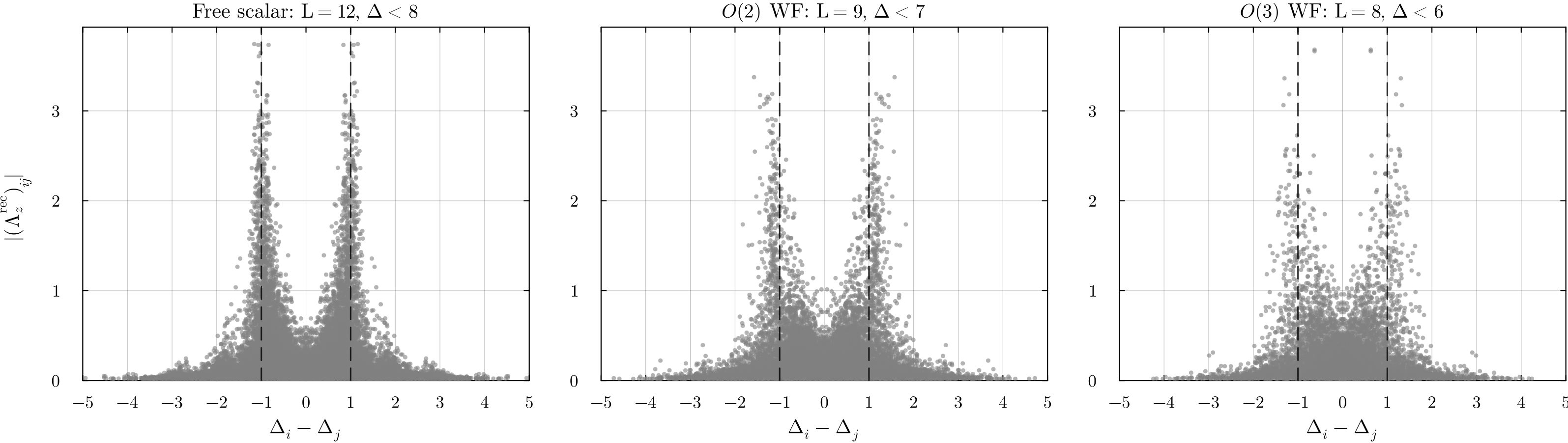}
    \caption{\justifying Matrix elements of the reconstructed conformal boost $\Lambda_z^{\rm rec}$ in the $D_{\rm rec}$ eigenbasis as a function of the energy difference $\Delta_i - \Delta_j$. 
    Left: Free scalar at $L=12$ for $\Delta < 8$.
    Middle: $O(2)$ Wilson-Fisher (WF) at $L=9$ for $\Delta < 7$.
    Right: $O(3)$ Wilson-Fisher at $L=8$ for $\Delta < 6$. 
    Subregions are chosen as $(A_1,A_2,A_3) = (\left\lfloor L/10 \right \rceil, \left\lfloor L/4 \right\rceil, \left\lfloor 2L/5 \right\rceil)$, where $\left\lfloor \ldots \right\rceil$ denotes the nearest integer.
    Only matrix elements with magnitude greater than $10^{-2}$ are shown. 
    Dashed vertical lines indicate the CFT expectations $\Delta_i - \Delta_j = \pm 1$.
    }
    \label{fig:Kz-matrix-elements-other-CFTs}
\end{figure}

\begin{figure}[!ht]
    \centering
    \includegraphics[width=\linewidth]{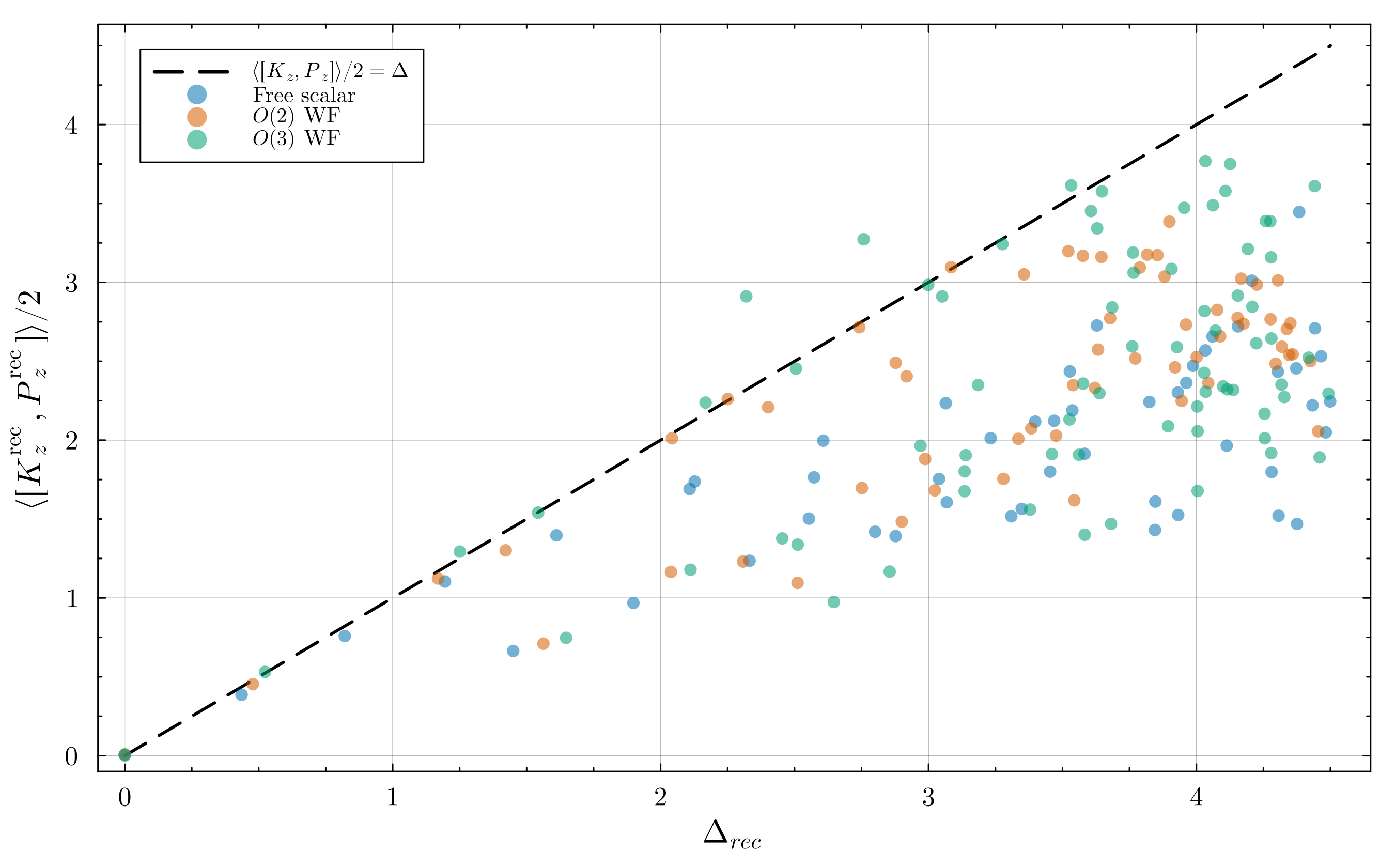}
    \caption{\justifying Expectation value of the commutator $[K_z^{\rm rec},P_z^{\rm rec}]/2$ for eigenstates of $D_{\rm rec}$ as function of their energy $\Delta_{\rm rec}$ for various approximate CFT groundstates. 
    Blue, orange, and green correspond to the free scalar at $L=12$, the $O(2)$ Wilson-Fisher CFT at $L=9$, and the $O(3)$ Wilson-Fisher CFT at $L=8$, respectively. 
    Dashed diagonal line denotes the CFT expectation.
    } 
    \label{fig:Other-CFTs-commutator}
\end{figure}

We first examine the reconstructed conformal boost operator $\Lambda_z^{\rm rec}$ in the $D_{\rm rec}$ eigenbasis. 
Figure~\ref{fig:Kz-matrix-elements-other-CFTs} shows that the largest matrix elements are concentrated near the expected energy differences $\Delta_i-\Delta_j = \pm 1$ for all three models, though the spread around these values varies between them.
We also test the commutation relation $[K_z,P_z] = 2D$ in Figure~\ref{fig:Other-CFTs-commutator} which shows that the low-energy states approximately satisfy this relation, while larger deviations appear at higher energies. 
The agreement varies among the models, which are studied at different system sizes (because they have different local Hilbert space dimensions).

\begin{figure}[!ht]
    \centering
    \includegraphics[width=\linewidth]{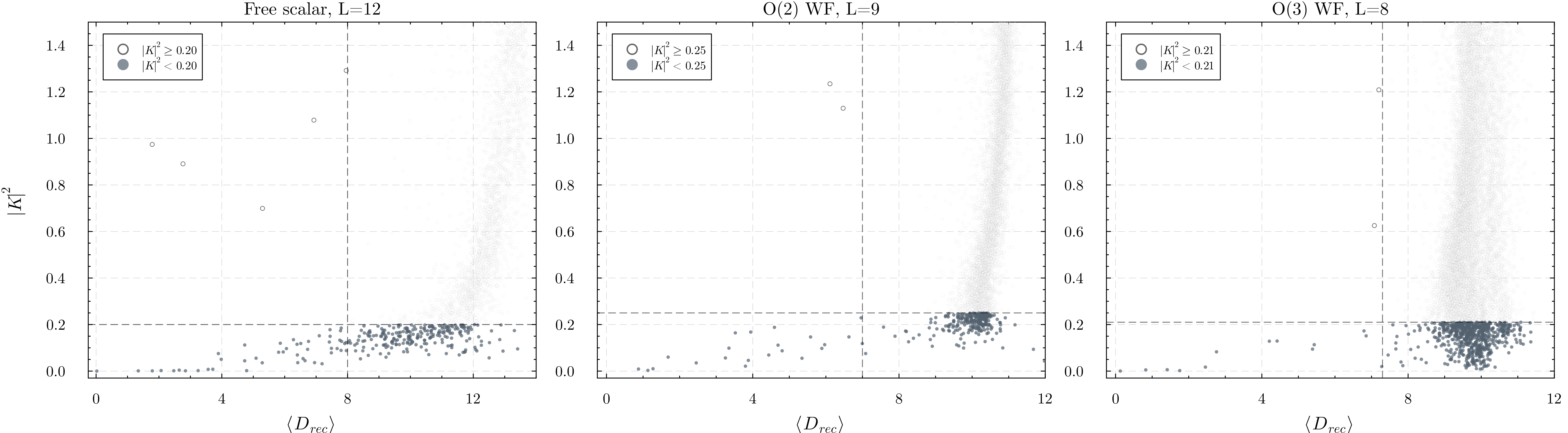}
    \caption{\justifying Low-lying eigenvalues of $|K|^2$ as a function of $\langle D_{\rm rec} \rangle$ for the free scalar at $L=12$ (left), the $O(2)$ Wilson-Fisher at $L=9$ (middle), and the $O(3)$ Wilson-Fisher at $L=8$ (right). 
    Expectation values of $D_{\rm rec}$ are evaluated in eigenstates of $|K|^2$.
    The opacity indicates whether the eigenvalue lies below the cutoff $\kappa_c$. 
    Vertical dashed line shows the estimated $\langle D_{\rm rec} \rangle$ above which the gap in the spectrum of $|K|^2$ starts closing.
    }
    \label{fig:other-CFTs-O1-gap}
\end{figure}

Next, we check whether the reconstructed conformal generators can be used to identify primary states.
We construct $|K|^2 = \sum_i (K_i^{\rm rec})^\dagger K_i^{\rm rec}$ and look for its approximate kernel.
Figure~\ref{fig:other-CFTs-O1-gap} shows that a set of low-$|K|^2$ states can be distinguished from the rest of the spectrum in all three models, particularly at low reconstructed energies $\langle D_{\rm rec} \rangle$.
As in the Ising model, we select the eigenstates below a cutoff $\kappa_c$ and rediagonalize $D_{\rm rec}$ within this subspace to obtain the reconstructed scaling dimensions. The dependence on $\kappa_c$ is then represented by the error bars in the following figures. 

\begin{figure}[!t]
    \centering
    \includegraphics[width=\linewidth]{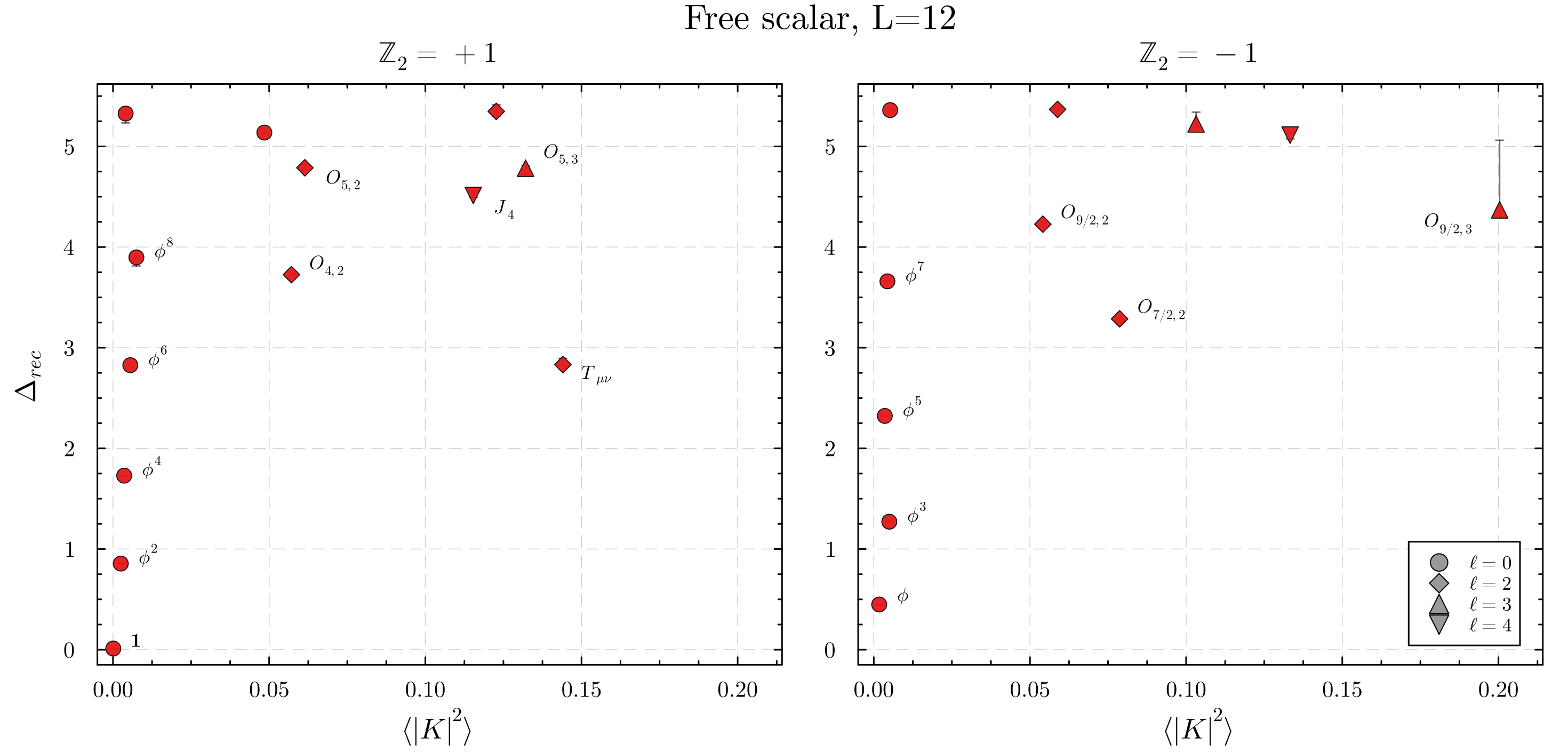}
    \caption{\justifying 
    Reconstructed low-$|K|^2$ spectrum of the free scalar CFT. We select eigenstates of $|K|^2$ with eigenvalues below $\kappa_c=0.35$ and rediagonalize $D_{\rm rec}$ within this subspace. 
    The resulting reconstructed scaling dimensions are plotted against $\langle |K|^2\rangle$, separated into the $\mathbb Z_2$-even (left) and $\mathbb Z_2$-odd (right) sectors. 
    Marker shapes indicate angular momentum $\ell$, and selected primary candidates are labeled. 
    The notation $O_{\Delta,\ell}$ denotes a primary candidate with expected CFT scaling dimension $\Delta$ and angular momentum $\ell$, while $J_4$ denotes the conserved spin-four current. 
    Vertical error bars represent the variation of the reconstructed scaling dimensions under changes in the cutoff over $\kappa_c\in[0.22,0.50]$.
    }
    \label{fig:primaries-FS-L12}
\end{figure}

We start by considering the free scalar CFT. 
Figure~\ref{fig:primaries-FS-L12} shows the reconstructed low-$|K|^2$ spectrum, separated into the $\mathbb{Z}_2$ even and odd sectors. 
We find the expected tower of scalar primaries $\phi^{2n}$ in the even sector and $\phi^{2n+1}$ in the odd sector, with scaling dimensions close to their CFT values $\Delta_{\phi^n} = n/2$.
Our results also show additional spinning primaries consistent with the free scalar spectrum, such as the energy-momentum tensor $T_{\mu \nu}$, a candidate for the conserved spin-four current $J_4$, and several additional spinning primary candidates.
Higher-energy states are consistent with the free scalar spectrum but exhibit larger finite-size deviations from the CFT values.

\begin{figure}[!ht]
    \centering
    \includegraphics[width=\linewidth]{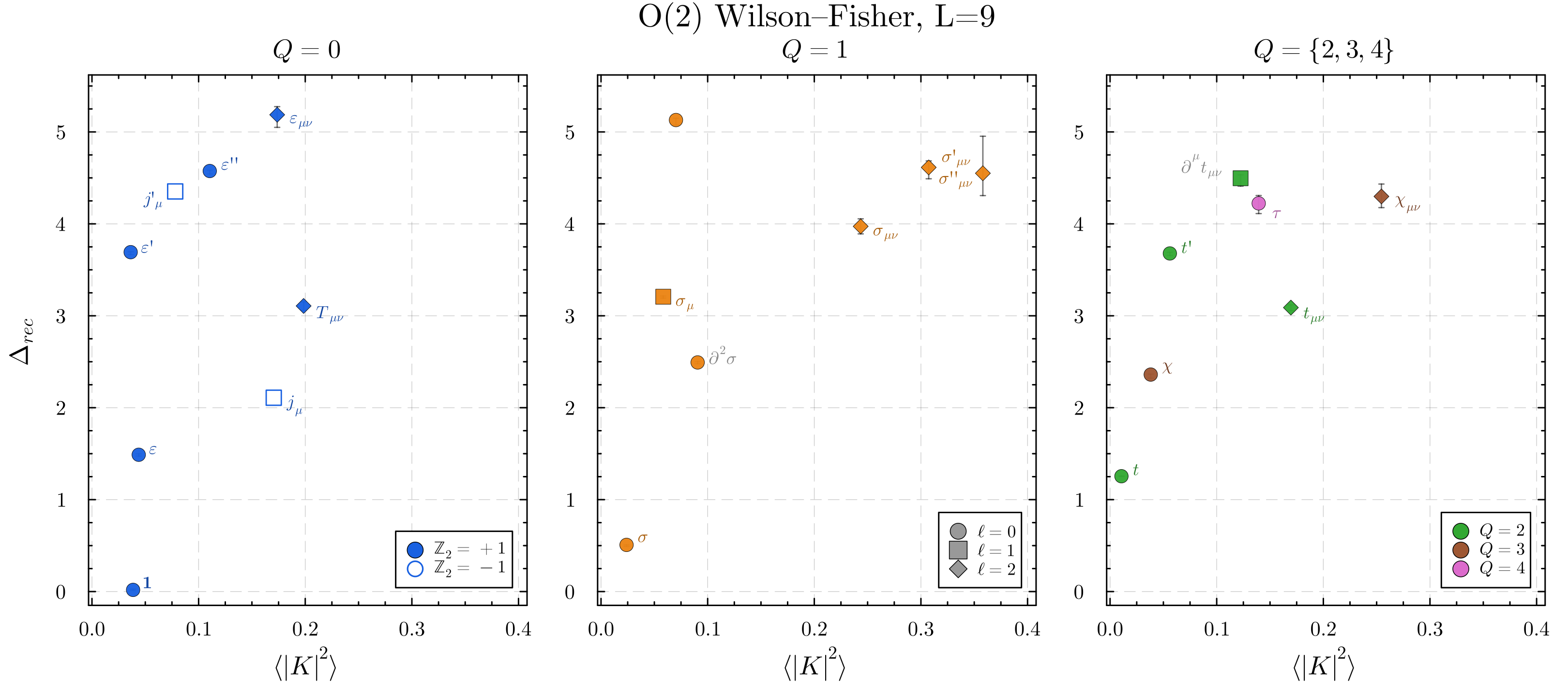}
    \caption{\justifying Reconstructed low-$|K|^2$ spectrum of the $O(2)$ Wilson-Fisher CFT. We select eigenstates of $|K|^2$ below $\kappa_c=0.50$ and rediagonalize $D_{\rm rec}$ within this subspace. The reconstructed scaling dimensions are plotted against $\langle |K|^2\rangle$. Left, middle, and right panels correspond to the charge sectors $Q=0$, $Q=1$, and $Q=2,3,4$, respectively. Marker shapes indicate angular momentum $\ell$, while selected primary candidates and low-$|K|^2$ descendants are labeled. Vertical error bars represent the variation of the reconstructed scaling dimensions under changes in the cutoff over $\kappa_c\in[0.42,0.80]$.
    }
    \label{fig:primaries-O2-L9}
\end{figure}

\begin{figure}[!ht]
    \centering
    \includegraphics[width=\linewidth]{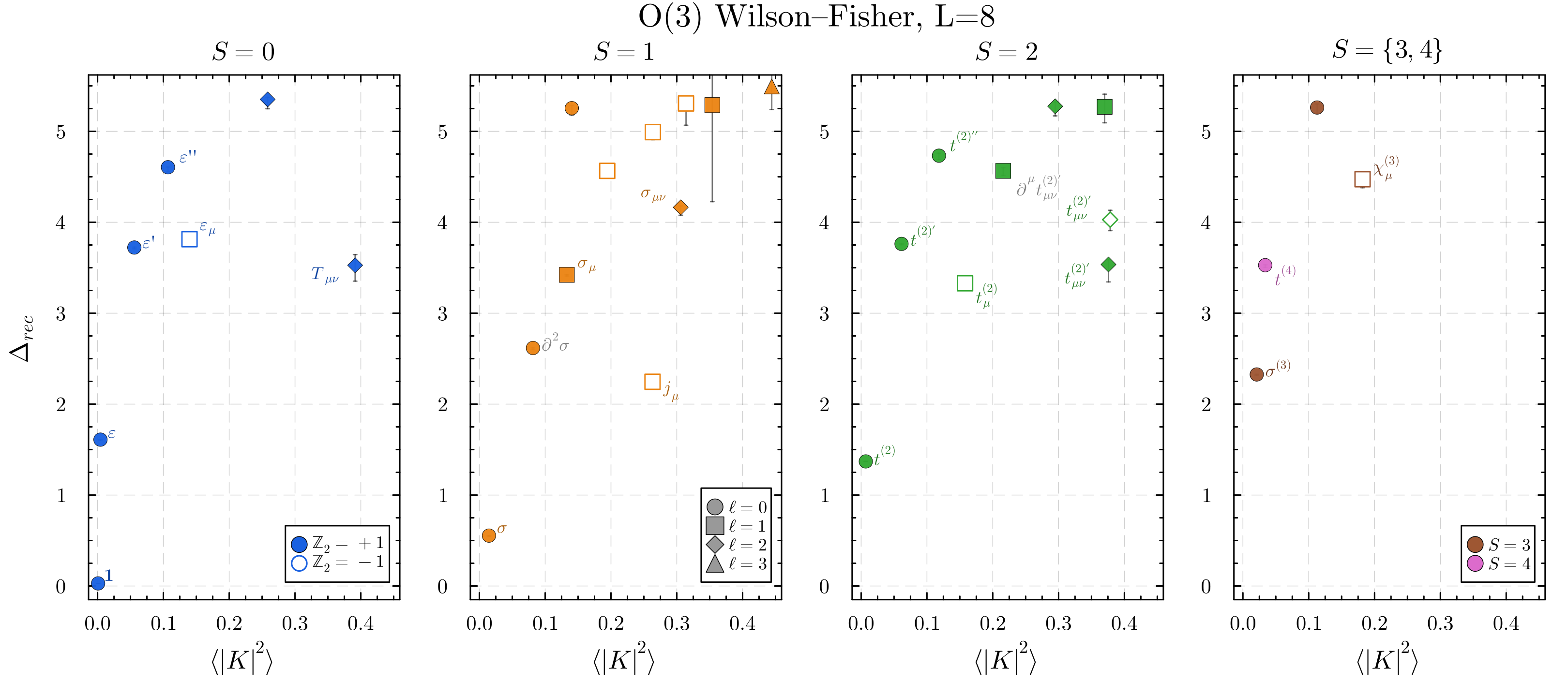}
    \caption{\justifying Reconstructed low-$|K|^2$ spectrum of the $O(3)$ Wilson-Fisher CFT. We select eigenstates of $|K|^2$ below $\kappa_c=0.55$ and rediagonalize $D_{\rm rec}$ within this subspace. The reconstructed scaling dimensions are plotted against $\langle |K|^2\rangle$. The four panels correspond to the internal-spin sectors $S=0$, $S=1$, $S=2$, and $S=3,4$, respectively. Marker shapes indicate spatial angular momentum $\ell$, while selected primary candidates and low-$|K|^2$ descendants are labeled. Vertical error bars represent the variation of the reconstructed scaling dimensions under changes in the cutoff over $\kappa_c\in[0.50,0.80]$.}
    \label{fig:primaries-O3-L8}
\end{figure}

Finally, Figures~\ref{fig:primaries-O2-L9} and~\ref{fig:primaries-O3-L8} show the reconstructed low-$|K|^2$ spectra of the $O(2)$ and $O(3)$ Wilson-Fisher CFTs, respectively. 
In both models, we recover the low-energy primaries $\sigma$, $\varepsilon$, $T_{\mu\nu}$ and the conserved current $j_\mu$, previously identified from the spectrum in~\cite{Camasca:2026awd}.
More interestingly, apart from these states, our results show good agreement with several higher-energy primaries recently reported in~\cite{dey2026:O2,dey2026:O3}. 
For $O(2)$, we find candidates for the neutral operators $\varepsilon''$, $j'_\mu$, and $\varepsilon_{\mu\nu}$, as well as spinning primaries in the charged sectors, such as $\sigma_\mu$, $\sigma_{\mu\nu}$, and $t_{\mu\nu}$, and higher-charge operators $\chi$ and $\chi_{\mu\nu}$. 
Similarly, for $O(3)$, we find a candidate for the parity-odd operator $\varepsilon_\mu$ in the $S=0$ sector, the spinning primaries $\sigma_\mu$ and $\sigma_{\mu\nu}$ in the $S=1$ sector, the higher scalar primaries $t^{(2)\prime}$ and $t^{(2)\prime\prime}$ in the $S=2$ sector, and the operators $\sigma^{(3)}$ and $t^{(4)}$ in the $S=3$ and $S=4$ sectors, respectively. 
In both models, we also find states consistent with almost-primary descendants, such as $\Box\sigma$ and divergences of nearly conserved spinning operators, which have small $|K|^2$, as expected from the exact CFT.

\section{Regularization with the icosahedron} \label{app:icosahedron}
We also applied the strategy of reconstructing the conformal generators on a lattice. There, instead of using the weighted integrals of entanglement Hamiltonians, we use a discrete sum. Explicitly, we considered the transverse-field Ising model at the critical point on the icosahedron studied in \cite{Lao:2023zis}, which is the 2+1D Ising CFT regulated on the lattice. 

On the icosahedron, there are three kinds of regions that we regard as a lattice version of ``round disk'', namely one site, a triangle of three sites, or the six sites in five triangles that share a vertex. The opening angle of each disk is computed by regarding the number of sites in it as the area of the disk. See \cite{Camasca:2026awd} for the detailed formula. The weighted sum is 
\begin{align}
    \overline{K}_{A,(l,m)} = \frac{1}{N_n}\sum_{|A| = n} Y_{lm}^*(\Omega_A) K_A,
\end{align}
where $n = 1, 3,6$, $\Omega_A$ is the angular coordinate of the center of the disk $A$, and $N_n$ is the number of $n$-site disks in the sum. We then follow the same formula to construct $\Lambda_z^{\rec}$. Using $D_{\rec}$ as described in \cite{Camasca:2026awd}, by taking the commutators, we can construct $K_z, P_z$. 

\begin{figure}[!ht]
    \centering
    \includegraphics[width=0.9\linewidth]{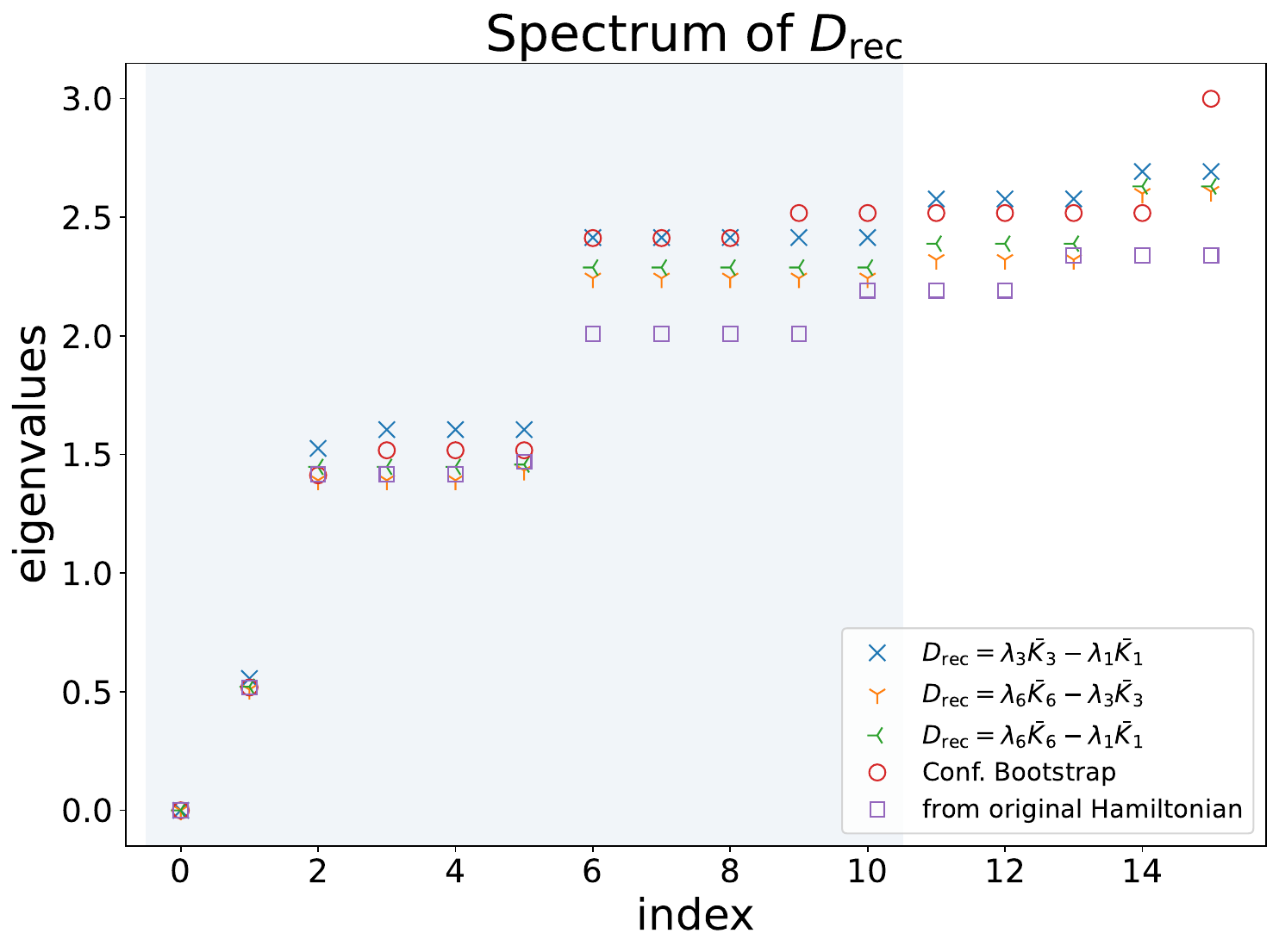}
    \caption{\justifying Reconstructed spectra on the icosahedron and their comparisons with the original Hamiltonian spectrum and the conformal bootstrap spectrum.}
    \label{fig:icosahedron-spectrum}
\end{figure}

Let us first review the quality of $D_{\rec}$ using entanglement Hamiltonians of two regions, shown in Fig.~\ref{fig:icosahedron-spectrum}. Among them, the best one is obtained using one-site and three-site entanglement Hamiltonians. Later, we shall focus on the reconstructed conformal generator in the subspace spanned by the first 11 eigenstates. 

\begin{figure}[!ht]
    \centering
    \includegraphics[width=0.9\linewidth]{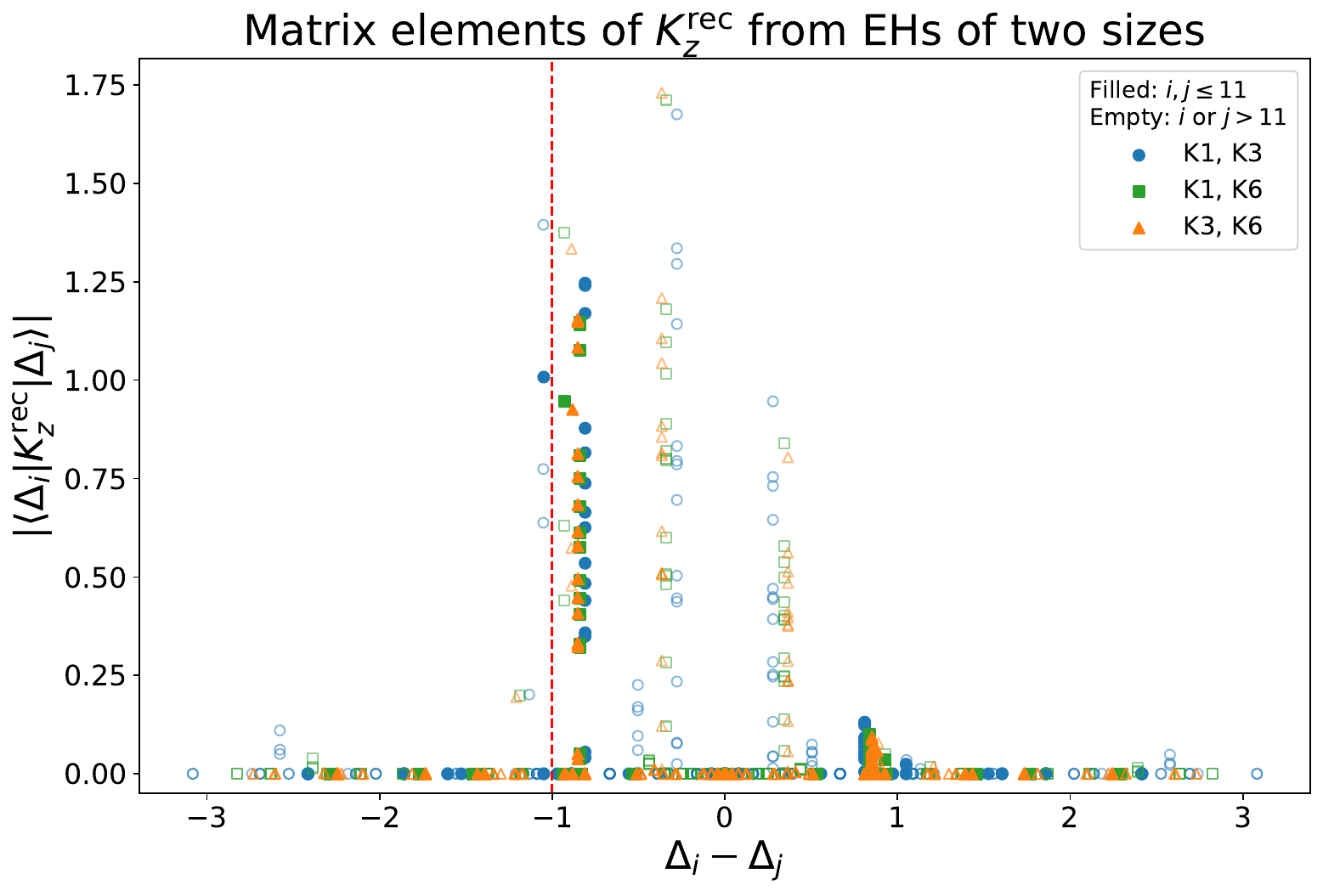}
    \caption{\justifying Matrix elements of $K_{z}^{\rec}$ in the eigenstates $\ket{\Delta_i}$ of $D_{\rec}$ on the icosahedron. $\Delta_i$ is the reconstructed scaling dimension, with $i$ labels the energy level.}
    \label{fig:icosahedron-matrix-elements}
\end{figure}

We then compute the matrix elements of $K_{z}^{\rec}$ in the eigenstates of $D_{\rec}$, shown in Fig.~\ref{fig:icosahedron-matrix-elements}. We can see that, among the first 11 levels, it roughly matches our expectation that the non-zero matrix elements shall accumulate at $\Delta_i - \Delta_j = -1$. Among the higher energy levels, the results are off the prediction from CFT, as the states are contaminated by UV degrees of freedom. 

\begin{figure}[!ht]
    \centering
    \includegraphics[width=0.9\linewidth]{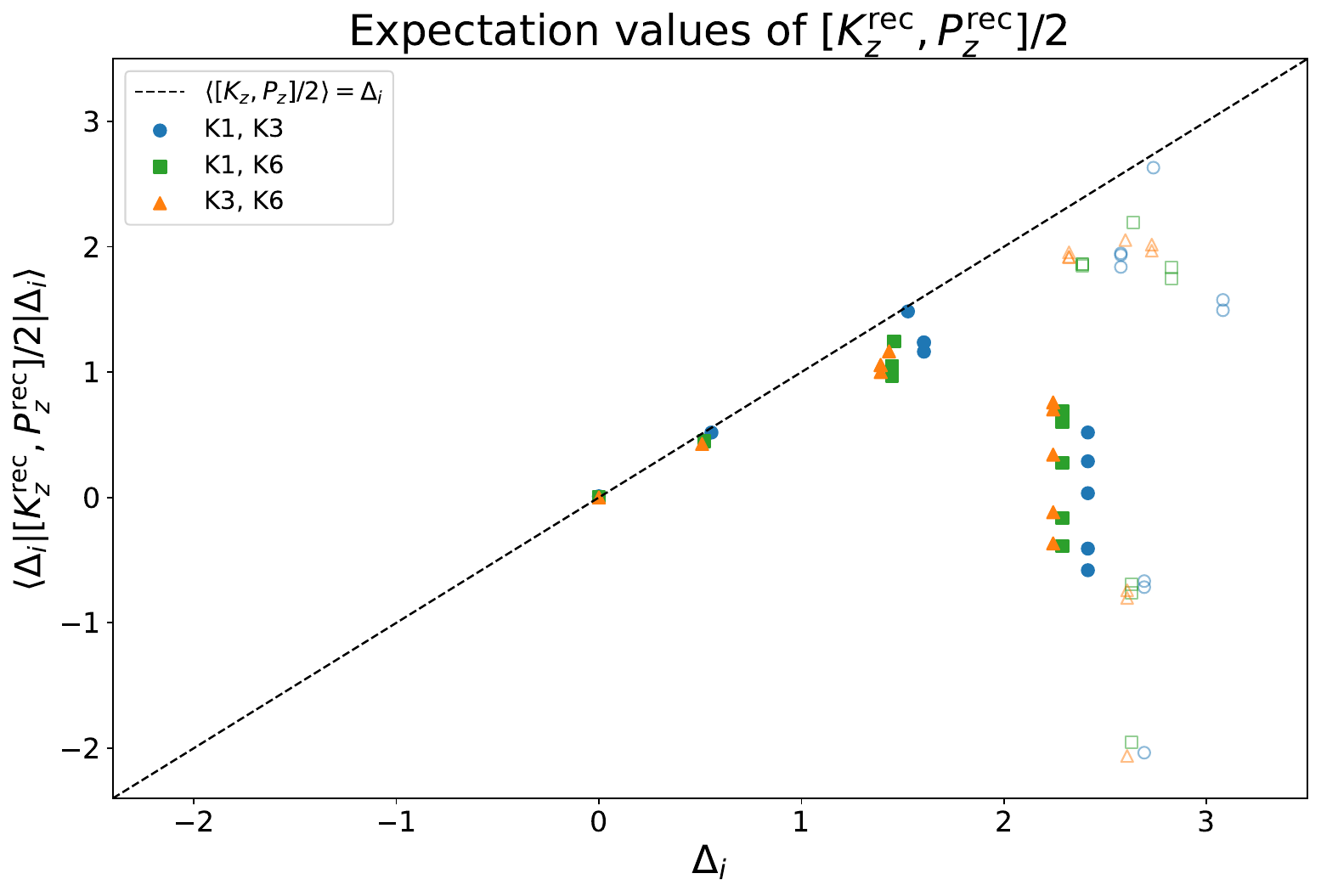}
    \caption{\justifying Expectation values of $[K_z^{\rec}, P_z^{\rec}]/2$ in the eigenstates $\ket{\Delta_i}$ of $D_{\rec}$ on the icosahedron with the reconstructed spectrum $\Delta_i$.}
    \label{fig:icosahedron-commutators}
\end{figure}

To test the algebraic relations, we compute $[K_z^{\rec}, P_z^{\rec}]/2$, which is expected to give $D_{\rec}$ if the state is a perfect CFT groundstate. Here we compute the expectation values of the commutators in the eigenstates of $D_{\rec}$. The results are shown in Fig.~\ref{fig:icosahedron-commutators}, from which we can see again that the low-lying states match the CFT prediction.  

\section{Entanglement bootstrap from trial CFT wavefunctions revisited}
In~\cite{Camasca:2026awd}, we studied the reconstructed spectra of the interesting analytic trial wavefunctions proposed in~\cite{He:2025ong} for the free scalar and Ising CFTs.
We found that the free scalar ansatz gives a spectrum consistent with the free scalar CFT, while the Ising ansatz produces a remarkably similar spectrum, despite its large overlap with the numerical Ising ground state.
Here, we use the reconstructed conformal generators to identify candidate primary states from these trial wavefunctions.

\begin{figure}[!ht]
    \centering
    \includegraphics[width=\linewidth]{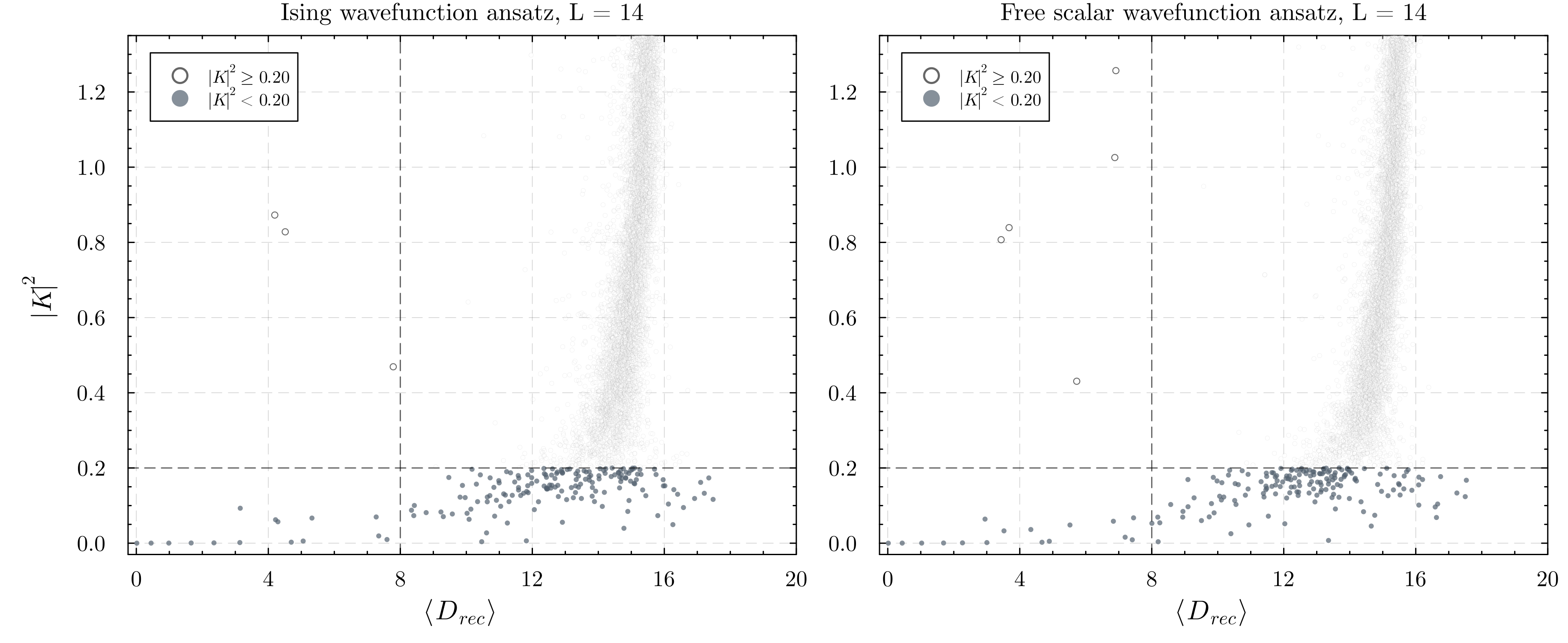}
    \caption{\justifying Low-lying eigenvalues of $|K|^2$ as a function of $\langle D_{\rm rec}\rangle$ for the Ising (left) and free scalar (right) trial wavefunctions at $L=14$. Expectation values of $D_{\rm rec}$ are evaluated in eigenstates of $|K|^2$.}
    \label{fig:gap-opening-ansatz}
\end{figure}

\begin{figure}[!ht]
    \centering
    \includegraphics[width=\linewidth]{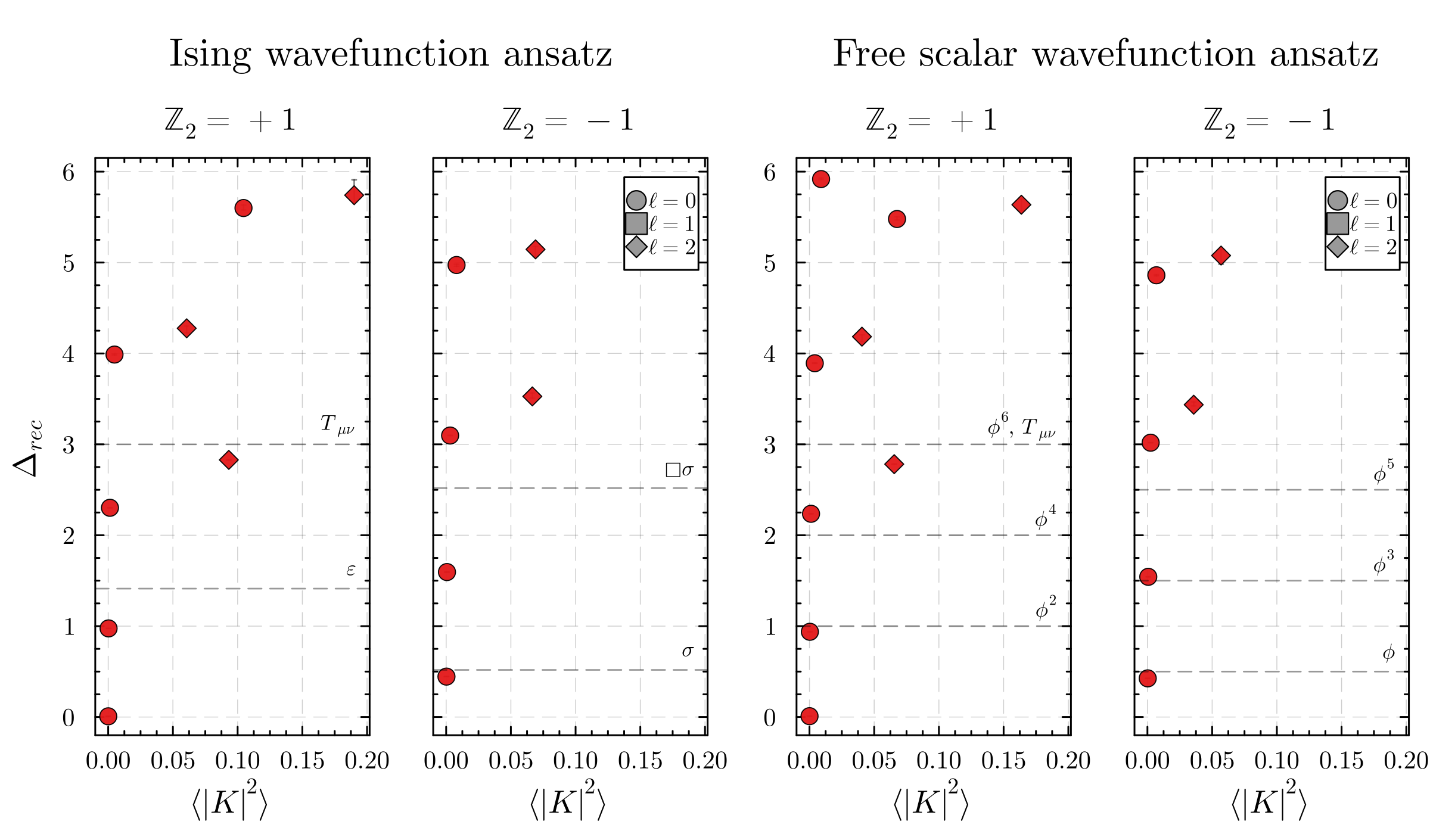}
    \caption{\justifying Reconstructed low-$|K|^2$ spectra of the Ising (left) and free scalar (right) trial wavefunctions at $L=14$. 
    We select eigenstates of $|K|^2$ below $\kappa_c=0.32$ and rediagonalize $D_{\rm rec}$ within this subspace. The resulting reconstructed scaling dimensions are plotted against $\langle|K|^2\rangle$, separated into the $\mathbb Z_2$-even and odd sectors. 
    Marker shapes indicate angular momentum $\ell$, while horizontal dashed lines denote the expected CFT scaling dimensions of the labeled operators. 
    Vertical error bars represent the variation of the reconstructed scaling dimensions under changes in the cutoff over $\kappa_c\in[0.22,0.50]$.}
    \label{fig:wavefunction-ansatz-primaries}
\end{figure}

Following the same procedure as in \S\ref{sec:extracting-primaries}, we extract candidate primary states by rediagonalizing $D_{\rm rec}$ within the low-$|K|^2$ subspace. 
Figure~\ref{fig:gap-opening-ansatz} shows the corresponding low-lying eigenvalues of $|K|^2$ while Figure~\ref{fig:wavefunction-ansatz-primaries} shows the resulting reconstructed spectra at $L=14$, separated into the $\mathbb Z_2$-even and odd sectors. 
We find that the similarity between the two ansätze persists in the low-$|K|^2$ spectra, particularly for the lowest scalar states.
In both cases, the first nontrivial scalar in the even sector has a reconstructed dimension close to the free-scalar value $\Delta_{\phi^2}=1$, rather than the Ising value $\Delta_\epsilon\simeq1.413$.
Similarly, in the odd sector, the Ising ansatz contains a low-$|K|^2$ scalar near $\Delta_{\rm rec}\simeq1.6$, consistent with the free-scalar primary $\phi^3$ with $\Delta_{\phi^3} = 3/2$.
The lowest reconstructed dimensions are also relatively insensitive under variations of the cutoff $\kappa_c$.
Our results provide further evidence that both trial wavefunctions exhibit low-energy candidate primary spectra resembling that of the free scalar CFT.

\newpage

\bibliographystyle{ucsd}
\bibliography{collection}
\end{document}

%% file: jm-tex-macros-public.tex
\usepackage{amssymb}
\usepackage{amsmath,bm}
\usepackage{amssymb}
\usepackage{graphicx}
\usepackage{amsfonts}         
\usepackage{fancybox}

\usepackage[usenames,dvipsnames]{xcolor}%

\usepackage[pagebackref,  %
bookmarks={false}, pdfauthor={John McGreevy}, pdftitle={Yay, physics!}]{hyperref}
\hypersetup{colorlinks=true, 
linkcolor=BrickRed, 
citecolor=Violet, 
filecolor=OliveGreen, 
urlcolor=RoyalBlue, 
filebordercolor={.8 .8 1}, 
urlbordercolor={.8 .8 0}}%
\usepackage{soul}%
\setstcolor{Red}

\definecolor{darkgreen}{rgb}{0,0.4,0}
\definecolor{darkred}{rgb}{0.4,0,0}
\definecolor{darkblue}{rgb}{0,0,0.4}
\definecolor{lightblue}{rgb}{.6,.6,0.9}

\definecolor{uglybrown}{rgb}{0.8,  0.7,  0.5}

\definecolor{palatinatepurple}{rgb}{0.41, 0.16, 0.38}
\definecolor{celebrationcolor}{rgb}{0.75,  0.0,  0.9}

\definecolor{Mathematica1}{rgb}{0.368417, 0.506779, 0.709798}
\definecolor{Mathematica2}{rgb}{0.880722, 0.611041, 0.142051}
\definecolor{Mathematica3}{rgb}{0.560181, 0.691569, 0.194885}

\definecolor{shadecolor}{rgb}{0.90,0.90,0.90}
\definecolor{DVcolor}{rgb}{0.95,  0.5,  0.2}
\definecolor{lightbluemuons}{rgb}{0.0,.65,1.0}

\definecolor{chartreuse}{rgb}{0.70, 1.00, 0.00}

\usepackage{tikz}
\usetikzlibrary{shapes}
\usetikzlibrary{trees}
\usetikzlibrary{snakes}
\usetikzlibrary{matrix,arrows} 				%
\usetikzlibrary{positioning}				%
\usetikzlibrary{calc,through}				%
\usetikzlibrary{decorations.pathreplacing}  %
\usepackage[tikz]{bclogo} 					%
\usepackage{pgffor}							%

\usetikzlibrary{decorations.markings}
\usetikzlibrary{intersections}				%

\usetikzlibrary{arrows,decorations.pathmorphing,backgrounds,positioning,fit,petri,automata,shadows,calendar,mindmap, graphs}
\usetikzlibrary{arrows.meta,bending}

\tikzset{
    vector/.style={decorate, decoration={snake}, draw},
    fermion/.style={postaction={decorate},
        decoration={markings,mark=at position .55 with {\arrow{>}}}},
    fermionbar/.style={draw, postaction={decorate},
        decoration={markings,mark=at position .55 with {\arrow{<}}}},
    fermionnoarrow/.style={},
    gluon/.style={decorate,
        decoration={coil,amplitude=4pt, segment length=5pt}},
    scalar/.style={dashed, postaction={decorate},
        decoration={markings,mark=at position .55 with {\arrow{>}}}},
    scalarbar/.style={dashed, postaction={decorate},
        decoration={markings,mark=at position .55 with {\arrow{<}}}},
    scalarnoarrow/.style={dashed,draw},
	vectorscalar/.style={loosely dotted,draw=black, postaction={decorate}},
}

\def\centerarc[#1](#2)(#3:#4:#5)%
    { \draw[#1] ($(#2)+({#5*cos(#3)},{#5*sin(#3)})$) arc (#3:#4:#5); }

\usepackage{enumitem}

\usepackage{slashed}

\usepackage[font=small,labelfont=bf]{caption}
\DeclareCaptionFont{tiny}{\tiny}
\usepackage{epstopdf}
\usepackage{wasysym}

\usepackage[yyyymmdd,hhmmss]{datetime}   
\usepackage{framed}  %
 \usepackage{mdframed}
 \usepackage{wrapfig}
 \usepackage{yfonts}  

\newmdenv[%
        backgroundcolor=lightgray,
    linecolor=black,
    outerlinewidth=2pt,
]{boxedandshaded}

\numberwithin{equation}{section}

\renewcommand{\theequation}{\arabic{section}.\arabic{equation}}

\newlength{\extraspace}
\newlength{\extraspaces}
\def\be{\begin{equation}}
\def\ee{\end{equation}}

\newcommand{\bea}{\begin{eqnarray}}
\newcommand{\eea}{\end{eqnarray}}

\def\eps{\epsilon}

\def\bra#1{\left\langle#1\right|}
\def\ket#1{\left|#1\right\rangle}

\def\II{\relax{I\kern-.10em I}}

\def\IB{\relax{\rm I\kern-.18em B}}

\def\ID{\relax{\rm I\kern-.18em D}}
\def\IE{\relax{\rm I\kern-.18em E}}
\def\IF{\relax{\rm I\kern-.18em F}}
\def\IG{\relax\hbox{$\inbar\kern-.3em{\rm G}$}}
\def\IGa{\relax\hbox{${\rm I}\kern-.18em\Gamma$}}
\def\IH{\relax{\rm I\kern-.18em H}}
\def\II{\relax{\rm I\kern-.18em I}}
\def\IK{\relax{\rm I\kern-.18em K}}

\def\inbar{\,\vrule height1.5ex width.4pt depth0pt}

\def\simgt{\hskip0.05in\relax{ 
\raise3.0pt\hbox{ $>$
{\lower5.0pt\hbox{\kern-1.05em $\sim$}} }} \hskip0.05in}

\def\lp10{\ell_p^{10}}
\def\lp11{\ell_p^{11}}
\def\R11{R_{11}}

\def\frac#1#2{{#1 \over #2}}

\def\norm#1{\vrule\hskip-1.2pt \vrule \hskip1.2pt #1  \hskip1.2pt \vrule\hskip-1.2pt \vrule}
\def\norm#1{|\hskip-1.5pt | \hskip1.5pt #1  \hskip1.5pt |\hskip-1.5pt |}

\def\norm#1{\vrule\hskip-1.2pt \vrule \hskip1.2pt #1  \hskip1.2pt \vrule\hskip-1.2pt \vrule}
\def\norm#1{|\hskip-1.5pt | \hskip1.5pt #1  \hskip1.5pt |\hskip-1.5pt |}
\newdimen\tableauside\tableauside=1.0ex
\newdimen\tableaurule\tableaurule=0.4pt
\newdimen\tableaustep
\def\phantomhrule#1{\hbox{\vbox to0pt{\hrule height\tableaurule width#1\vss}}}
\def\phantomvrule#1{\vbox{\hbox to0pt{\vrule width\tableaurule height#1\hss}}}
\def\sqr{\vbox{%
  \phantomhrule\tableaustep
  \hbox{\phantomvrule\tableaustep\kern\tableaustep\phantomvrule\tableaustep}%
  \hbox{\vbox{\phantomhrule\tableauside}\kern-\tableaurule}}}
\def\squares#1{\hbox{\count0=#1\noindent\loop\sqr
  \advance\count0 by-1 \ifnum\count0>0\repeat}}
\def\tableau#1{\vcenter{\offinterlineskip
  \tableaustep=\tableauside\advance\tableaustep by-\tableaurule
  \kern\normallineskip\hbox
    {\kern\normallineskip\vbox
      {\gettableau#1 0 }%
     \kern\normallineskip\kern\tableaurule}%
  \kern\normallineskip\kern\tableaurule}}
\def\gettableau#1 {\ifnum#1=0\let\next=\null\else
  \squares{#1}\let\next=\gettableau\fi\next}

\def\({\left(}
\def\){\right)}

\def\lsim{\mathrel{\mathstrut\smash{\ooalign{\raise2.5pt\hbox{$<$}\cr\lower2.5pt\hbox{$\sim$}}}}}
\def\gsim{\mathrel{\mathstrut\smash{\ooalign{\raise2.5pt\hbox{$>$}\cr\lower2.5pt\hbox{$\sim$}}}}}

\def\overleftrightarrow#1{\vbox{\ialign{##\crcr
     $\leftrightarrow$\crcr\noalign{\kern-0pt\nointerlineskip}
     $\hfil\displaystyle{#1}\hfil$\crcr}}}
     
     \def\overleftarrow#1{\vbox{\ialign{##\crcr
     $\leftarrow$\crcr\noalign{\kern-0pt\nointerlineskip}
     $\hfil\displaystyle{#1}\hfil$\crcr}}}

\def\ie{{\it i.e.}}

\def\gSO{\textsf{SO}}

\newif{\ifeq}           %
\eqtrue                 %

\newcounter{lecturecounter}

%% file: macros.tex
\newcommand{\interior}[1]{%
  {\kern0pt#1}^{\mathrm{o}}%
}

\def\={\phantom{{}={}}}